\documentclass[twocolumn,showpacs,pre,eqsecnum,citeautoscript,amsmath,amssymb,floatfix,superscriptaddress]{revtex4-1}
\usepackage{bm,color,amsmath,amssymb,mathrsfs,latexsym,graphicx,psfrag,float}

\newcommand{\opt}[1]

\usepackage{comment}
\usepackage{hyperref}
\usepackage{enumerate}
\usepackage{txfonts}
\usepackage{bbm}
\usepackage{ifthen}
\usepackage{cleveref}
\usepackage{dsfont}
\usepackage{tabularx}
\newcolumntype{L}[1]{>{\raggedright\arraybackslash}p{#1}} 
\newcolumntype{C}[1]{>{\centering\arraybackslash}p{#1}} 
\newcolumntype{R}[1]{>{\raggedleft\arraybackslash}p{#1}} 

\newcommand{\eq}[2]{\begin{align}\label{#1} #2 \end{align}}

\newcommand{\cre}[2]{\bar{#1}_{#2}}

\newcommand{\ann}[2]{#1_{#2}}
\newcommand{\sub}[1]{_{\mbox{\tiny #1}}}

\newcommand{\creo}[2]{#1^{\dagger}_{#2}}
\newcommand{\anno}[2]{#1^{\phantom{\dagger}}_{#2}}
 
\usepackage{color}
\newcommand{\changed}[1]{{\color{black}#1}}
\newcommand{\mh}[1]{{\color{black}#1}}

\begin{document}
\title{Prethermalization and Thermalization of a Quenched Interacting Luttinger Liquid}

\author{Michael Buchhold}
\affiliation{Institut f\"ur Theoretische Physik, Universit\"at zu K\"oln, D-50937 Cologne, Germany}
\author{Markus Heyl}
\affiliation{Physik Department, Technische Universit\"at M\"unchen, 85747
Garching, Germany}
\author{Sebastian Diehl}
\affiliation{Institut f\"ur Theoretische Physik, Universit\"at zu K\"oln, D-50937 Cologne, Germany}
\begin{abstract}
We study the relaxation dynamics of interacting, one-dimensional fermions with band curvature after a weak quench in the interaction parameter at zero temperature. Our model lies within the class of interacting Luttinger Liquids, where the harmonic Luttinger theory is extended by a weak integrability breaking phonon scattering term. In order to solve for the non-equilibrium time evolution, we use quantum kinetic equations exploiting the resonant but subleading character of the phonon interaction term. The interplay between phonon scattering and the quadratic Luttinger theory leads to the emergence of three distinct spatio-temporal regimes for the fermionic real-space correlation function. It features the crossover from a prequench to a prethermal state, finally evolving towards a thermal state on increasing length and time scales. The characteristic algebraically decaying real-space correlations in the prethermalized regime become modulated by an amplitude, which is decaying in time according to a stretched-exponential as an effect of the interactions. The asymptotic thermalization dynamics is governed by energy transport over large distances from the thermalized to the non-thermalized regions via macroscopic, dynamical slow modes. This is revealed in an algebraic decay of the system's effective temperature. The numerical value of the associated exponent agrees with the dynamical critical exponent of the Kardar-Parisi-Zhang universality class. We also discuss a criterion for the applicability of this theory away from the integrable limit of non-interacting fermions.
 \end{abstract}
\maketitle

\section{Introduction}
It is a property of fundamental importance within statistical physics that generic and realistic thermodynamic systems exhibit one particular state -- thermal equilibrium -- which is always approached, irrespective of the initial condition. Yet the important question of which
microscopic conditions are necessary or sufficient for the thermalization of a closed quantum many-body system is still largely unanswered~\cite{Polkovnikov11}. This is of particular importance, especially because there exists a specific class of isolated quantum systems, termed integrable, for which relaxation to thermal states is prevented due to the presence of an extensive number of (quasi-)local conservation laws~\cite{Rigol06,Rigol07,Kollar11}.
Such particular systems often represent isolated points in the parameter space of physical many-body systems and demand a precise tuning of the microscopic parameters. Nevertheless, these models are very valuable because they often represent fixed points of renormalization group theories and as such contain the low-temperature equilibrium properties of a much wider class of systems. This directly leads to an apparent dilemma in quantum many-body theory which has attracted a lot of interest recently. In particular, beyond equilibrium these integrable models become nongeneric as they fail to thermalize. Instead, they are trapped in extended prethermal states described by nonthermal generalized Gibbs ensembles~\cite{Polkovnikov11,Rigol06,Rigol07,Rigol08,Kollar11,cazalilla06,cardycalabrese06,Barthel08}. Resolving this dilemma is one of the major challenges for the understanding of the coherent dynamics of quantum many-body systems.

In this work we address this question for a paradigmatic low-energy model: the Luttinger liquid~\cite{haldane81,haldane81b,giamarchi04}, representing the fixed point theory of systems of interacting fermionic particles in one dimension at low temperatures. The Luttinger liquid is an integrable theory failing to thermalize but rather exhibiting a description in terms of a generalized Gibbs ensemble~\cite{cazalilla06,Iucci09,Iucci10}. Here, we will be interested in the nonequilibrium dynamics in the presence of a weak fermionic band curvature, which represents a generic perturbation, irrelevant in the low-energy equilibrium limit, but relevant on intermediate to long time scales in order to drive the crossover towards thermalization. 

The increasing number of cold atom experiments performed under out of equilibrium conditions~\cite{Greiner2002ux,kinoshita,Hofferberth06,trupke13,schmiedmayer12,nagerl13,meinert14,preiss15,hild14,cheneau12} has driven significant interest in the theoretical understanding of the non-equilibrium dynamics in quantum many-body systems. Importantly, these experiments share a remarkable isolation from the environment, thereby probing the purely coherent unitary time evolution on the experimentally relevant time scales. This has paved the way to experimentally study the constrained relaxational dynamics of quantum systems close to integrability~\cite{kinoshita,Langen14,agarwal14,schmiedmayernphys12}, showing unconventional properties due to the anticipated (quasi-)local conservation laws.  Although the inherent integrability breaking terms, resulting from, e.g., imperfections in the particle-particle interactions or higher orbital modes, are considered to be weak, they are believed to eventually cause relaxation to thermal states on long-time scales. Yet a full understanding of this process has not been achieved so far.
Within the current understanding, however, the thermalization dynamics of quantum many-body systems with weak integrability-breaking perturbations is expected to occur via a two-stage process. Initially, the dynamics of local observables at transient and intermediate time scales are controlled by the corresponding integrable theory
serving as a metastable attractor for the non-integrable dynamics~\cite{Moeckel,Kollar11,Stark13}. 
This trapping in a metastable state has been termed prethermalization~\cite{berges_pretherm,Moeckel} and is expected to exist for several non-integrable models and models close to integrability~\cite{Moeckel,Eckstein09,Essler14,Kollar11,Rosch08,Marcuzzi13,Nessi15,Fagotti14,Fagotti15,Babadi2015,Bertini15}. 
In the quasi-particle picture, prethermalization is associated with the initial formation of well-defined excitations \cite{Moeckel} which leads to a dephasing of all terms that are not diagonal in quasi-particle modes, i.e. to a projection of the initial density matrix onto the diagonal ensemble in the quasi-particle basis. After this intermediate  quasi-particle formation, the dynamics eventually  crosses over to the thermalization regime, where weak  quasi-particle scattering leads to a slow redistribution of energy and establishes detailed balance between the different modes. This causes asymptotic thermalization on long time scales compatible with the Eigenstate-Thermalization-Hypothesis~\cite{Srednicki94,Deutsch91,Rigol08,Gibbs,Biroli10}.

In equilibrium, the fermionic band curvature in the Luttinger liquid, because irrelevant in the renormalization group sense,  does not modify static correlation functions, which are well described by the quadratic Luttinger theory. Importantly, however, the curvature has a strong impact on frequency-resolved fermionic quantities. This has been observed in Coulomb drag experiments \cite{Debray01,Debray02}, which could not be explained in terms of a quadratic Luttinger theory. In a hydrodynamic representation, the band curvature describes resonant scattering processes between the elementary phononic excitations of the system, such that perturbation theory is plagued by divergences due to the resonant nature of the interactions. Important first approaches to the interacting Luttinger liquid applied a self-consistent Born approximation in order to determine the phonon self-energy on the mass-shell \cite{andreev80,samokhin98,zwerger06}. However, these works were unable to explain the frequency-dependence of the self-energy, which appeared to be non-negligible for dynamic observables. Using a combination of bosonization and subsequent refermionization a general theory has been developed which has been very successful in determining spectral equilibrium properties such as the dynamic structure factor and the  fermionic spectral function in thermal equilibrium \cite{pustilnik03,pustilnik07,imambekov09,imambekov09a}. Importantly for the scope of the present work, however, it has not yet been possible to generalize this methodology to systems out of equilibrium. Only recently, these equilibrium results have been recovered by a quantum hydrodynamic approach \cite{lamacraft15,lamacraft15a}, showing that hydrodynamics is also capable of controlling the resonant phonon interactions.

The theoretical finding of these works is that the elementary excitations are no longer described in terms of bosonic quasi-particles with exact energy-momentum relation $\omega=u|q|$ but dissolve into a continuum of excitations. This continuum, however, is energetically confined between two well-defined excitation branches $\epsilon^-_q<\omega<\epsilon^+_q$ (with $\epsilon^{\pm}_q\rightarrow 0$ as $q\rightarrow0$) at which the spectral weight of the bosonic excitations features algebraic divergences, reflected in corresponding divergences of the dynamical structure factor. This fine structure in the bosonic spectral weight, and equivalently self-energy, makes the development of a general kinetic theory for \emph{frequency-resolved} observables a very demanding task, which has not yet found a satisfactory solution. However, as will be shown in this work, static properties and their time evolution are nevertheless accessible.

The goal of this work is to study \changed{the escape out of the prethermalization regime and the crossover towards thermalization} in Luttinger liquids with \changed{quadratic} fermionic dispersion on the basis of a hydrodynamic description. Specifically, we aim at formulating a kinetic theory for the momentum distribution of the phononic degrees of freedom \changed{taking into account the leading nonlinear corrections due to the quadratic dispersion. While in this way we are able to describe the escape out of the prethermalization regime in a controlled way, the final asymptotic thermalization of the system might be modified by the more subleading, off-resonant contributions which we do not consider here.} 
The kinetic equation describes the time-evolution of the phonon momentum distribution and is
suitable in the long-wavelength limit and for weak quenches but still goes beyond the regime of linear response. In turn this kinetic theory gives a valid description  for the fermionic occupation distribution in the vicinity of the Fermi points where the anticipated fine-structure of the bosonic spectral weight only gives subleading contributions.
This "semi-static" -- and as a consequence tractable -- description, covers the forward time evolution of any static, i.e. frequency independent, observable.
We show that the dynamics of precisely these frequency independent observables depend only on the time evolution of the momentum distribution of excitations $n_q$ and can be captured within a kinetic theory. 
The justification for this approach is the subleading width of the excitation spectrum $|\epsilon^+_q-\epsilon^-_q|\ll u|q|$ compared to the phonon energy for all relevant $q$ (below the Luttinger liquid cutoff), which is equivalent to the statement that even in the presence of the non-linearity the continuum of excitations in the hydrodynamic description is tightly bound to the mass-shell. This condition replaces the common quasi-particle criterion \cite{kamenevbook} and enables a thorough kinetic description.

\mh{The applicability of the kinetic equation requires the preformation of well-defined quasi-particles out of the bare particles which occurs during the process of prethermalization before the quasi-particle scattering sets in. We, however, find that close to the integrable point of vanishing fermionic interactions quasi-particle formation becomes very slow shifting the applicability of the theory for weakly interacting fermions to long time scales and far distances. We give quantitative estimates of the corresponding spatio-temporal scales of the breakdown of the kinetic theory. Not too close to the noninteracting point, however, the kinetic equation is well justified and allows us to study the escape out of the prethermalization regime towards thermalization.}
In the regime of applicability, the kinetic equation leads in the asymptotic long-time limit to a linearized quantum Boltzmann equation whose attractor is the desired thermal Gibbs state.
We find that the thermalization dynamics out of the prethermal state is triggered by short wavelength modes and afterwards progressing algebraically slowly towards longer wavelengths. 
Whether this is a generic feature of weakly-perturbed integrable theories, is an important and interesting question for future work. 

\begin{figure}
\centering
  \includegraphics[width=1\linewidth]{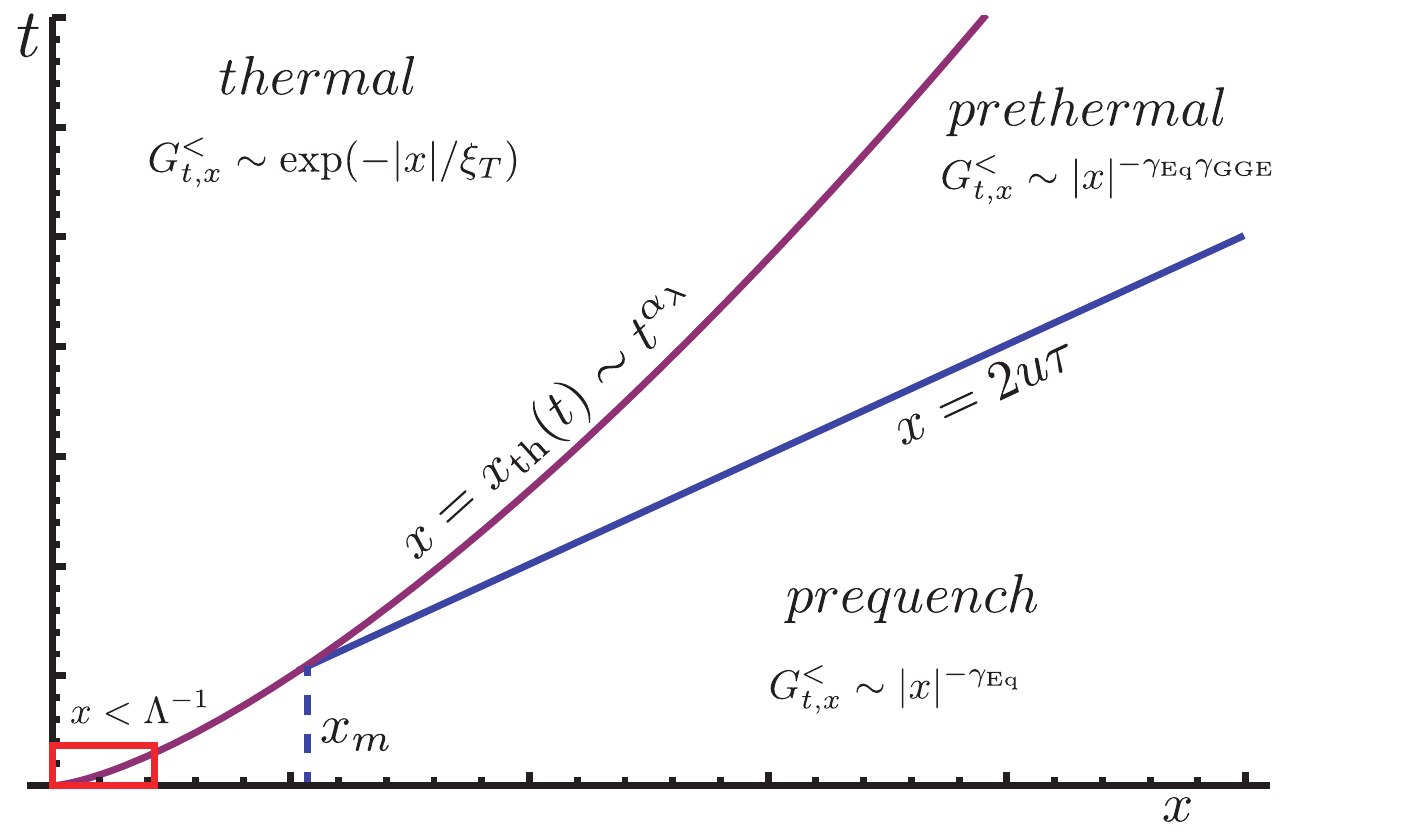}
  \caption{Illustration of the spatio-temporal thermalization and prethermalization dynamics in terms of the fermionic Green's function $G^<_{t,x}$. For long distances, $x>2ut, x>x\sub{th}$, the Green's function is determined by the quasi-particles of the initial state and feature algebraic decay in real-space corresponding to the pre-quench state of the system, modulated by a amplitude decaying as a stretched exponential in time. In the intermediate regime $2ut<x<x\sub{th}(t)$, the corresponding quasi-particles correspond to the post-quench Hamiltonian but are distributed according to a non-equilibrium distribution function, inducing a prethermal real-space scaling behavior $|G^<_{t,x}|\sim |x|^{-\gamma\sub{Eq}\gamma\sub{GGE}}$. On short distances $x<x\sub{th}(t)$, the Green's function is thermal, $\sim \exp{-|x|/\xi_T}$, described by an effective temperature $\tilde{T}_t$ and a corresponding thermal correlation length $\xi_T$. 
The scaling $x\sub{th}(t)\sim t^{\alpha_{\lambda}}$, $\alpha<1$ implies that there exists a minimal distance $x_m$, for which no clear prethermal regime can be identified since the scattering of quasi-particles is equally fast than the formation of quasi-particles. In this regime, the kinetic theory can not be applied. The short distance regime $x<\Lambda^{-1}$, for which Luttinger theory is invalid, occupies a negligibly small short time regime.}
  \label{fig:QuenchDiag}
\end{figure}

The main result of this work is a spatio-temporal decomposition of correlations in the studied nonlinear Luttinger Liquid, which is illustrated in Fig.~\ref{fig:QuenchDiag}. 
By analyzing the equal-time fermionic Green's function $G\sub{t,x}^{<}$, the Fourier transform of the fermionic occupation distribution, we find three regimes which we term prequench, prethermal, and thermal and which are separated by two crossover scales $x\sub{th}(t)$ and $x\sub{pt}(t)$ obeying $x\sub{th}(t) < x\sub{pt}(t)$. The crossover scale $x\sub{pt}(t)=2ut$ sets the light cone~\cite{cardycalabrese06} with $u$ the sound velocity of the elementary bosonic excitations of the integrable theory. Causality implies that for distances $x \gg x\sub{pt}(t)$ the system's properties are not yet influenced by the nonequilibrium protocol, but are rather given by the initial state yielding the notion of the prequench regime. Inside the light cone for distances $x<x\sub{pt}(t)$ we identify a further crossover scale $x\sub{th}(t)$ separating the prethermal and thermal spatial regions. For distances $x\sub{th}(t) \ll x \ll x\sub{pt}(t)$ the system's spatial correlations are controlled by the integrable theory which for long times are determined by the associated generalized Gibbs ensemble. This regime is therefore called prethermal. 
Interestingly, the thermalization dynamics, triggered by the weak fermionic nonlinearity, sets in at even smaller scales $x\ll x\sub{th}(t)$. At these distances, the correlations approach their thermal form. However, the associated effective temperature $\tilde T_t$ is larger than the expected temperature $T$ for the asymptotic fully thermalized state. Instead $\tilde T_t$ is a dynamical quantity approaching $T$ only algebraically slowly due to macroscopic dynamical slow modes.


Kinetic equations have been successfully applied to Luttinger liquids with a cosine potential, resulting from particle backscattering in Refs.~\cite{mitrarosch,Tavora13}. For Luttinger Liquids with cubic interactions a kinetic equation approach has been derived in Ref.~\cite{buchholdmethod}. The latter makes use of non-perturbative Dyson-Schwinger equations in order to solve the time-evolution of the phonon distribution function in the presence of the RG-irrelevant but resonant interactions. This kinetic equation approach is particularly well suited for Luttinger models close to the ground state, i.e. with a small number of phononic excitations but can also be applied to excited states, as long as the Luttinger criterion is satisfied locally, i.e. as long as the phonon density $n_q<\Lambda/|q|$ for all momenta $|q|$. Based on Ref.~\cite{buchholdmethod}, we can give explicit criteria for the validity of this approach for the fermionic dynamics after we have introduced the quench scenario.

This paper is organized as follows. We introduce the studied model system, the interacting Luttinger Liquid, in Sec.~\ref{sec:interactingluttingerliquid}. The main results are summarized in Sec.~\ref{sec:summaryofresults}. The derivation of the kinetic equations, which is used to solve the complex quantum many-body problem is presented in Sec.~\ref{sec:kinetic_equation}. It is analyzed and numerically  solved in Sec.~\ref{sec:thermalization_dynamics}, where we also give the derivation of the main results.

\section{Interacting Luttinger Liquid}
\label{sec:interactingluttingerliquid}
The simplest form of an interacting Luttinger Liquid emerges as the effective long-wavelength description of spinless interacting fermions with quadratic (i.e. dispersive) corrections to a perfectly linear dispersion around the Fermi energy \cite{haldane81,haldane81b,rozhkov05,Proto14a}. Although the fermionic band curvature is irrelevant in the sense of the renormalization group (RG)~\cite{rozhkov05} and does therefore not modify the static infrared behavior of the fermions, it is visible in dynamic observables, such as the fermionic spectral function or the dynamical structure factor \cite{imambekov08,zwerger06,affleck06,pustilnik06,Heyl2010,imambekov12}. In this work we will show that in a non-equilibrium situation, the quasi-particle scattering induced by the band curvature leads to a dynamical redistribution of energy and allows the system to relax towards a thermal state. Thus, the system becomes generic. This kind of relaxation is absent for non-dispersive fermions, since the corresponding model, the linear Luttinger model, is integrable. The fermionic band curvature breaks the integrability of the linear model and therefore, even though RG irrelevant, is the leading order term that drives the system away from a prethermal, i.e. GGE-type, dynamical fixed point and towards a thermal one.

The Luttinger liquid in its fermionic representation is described in terms of left and right moving spinless fermions (labeled with $\eta=\pm$), created and annihilated by operators $\creo{\psi}{\eta,x}, \anno{\psi}{\eta,x}$. The Hamiltonian is
\begin{equation}H=-\sum_{\eta}\int_x \creo{\psi}{\eta,x}\left(i\eta v\sub{F}\partial_x+\frac{1}{2m}\partial^2_x\right)\anno{\psi}{\eta,x}+\frac{1}{2}\int_{x,x'}g(x-x')\rho_x\rho_{x'},\label{Eq1}
\end{equation}
with the combined density $\rho_x=\rho_{+,x}+\rho_{-,x}=\creo{\psi}{+,x}\anno{\psi}{+,x}+\creo{\psi}{-,x}\anno{\psi}{-,x}$. The interaction, characterized by $g(x-x')$, is supposed to be short ranged in space (decaying faster than algebraic) but has a short distance cutoff of the order of the Luttinger cutoff $\Lambda^{-1}$. In the long-wavelength limit, particles with a wave-length larger than the effective range of the potential only experience a contact potential, $g(q)=g_0$, where $g_0$ is the interaction strength at zero momentum. In order to regularize the interaction in the ultraviolet (UV) regime, which is required to obtain a non-diverging, quench induced interaction energy, it is cut off at the UV scale $\Lambda$, i.e. $g(q)=g_0\theta(\Lambda-|q|)$. 

The bosonized version of the Hamiltonian in the absence of band curvature describes the well-known Luttinger model
\begin{equation}\label{Eq2}
H\sub{LL}=\int_x uK\left(\partial_x\theta_x\right)^2+\frac{u}{K}\left(\partial_x\phi_x\right)^2
\end{equation}
with sound velocity $u=\frac{v\sub{F}}{K}$ and Luttinger parameter $K=\left(1+\frac{g_0}{\pi v\sub{F}}\right)^{-\frac{1}{2}}$. In addition, due to the fermionic band curvature, a cubic nonlinearity occurs \cite{affleck06,zwerger06,aristov,imambekov08,lamacraft15}
\eq{Eq3}{
H\sub{NL}=\frac{1}{m}\int_x \left(\partial_x\theta_x\right)^2\partial_x\phi_x,}
such that the complete bosonized Hamiltonian is $H=H\sub{LL}+H\sub{NL}$. Due to the linear dispersion of the Luttinger quasi-particles, $H\sub{NL}$ describes scattering processes on a highly degenerate bosonic manifold, i.e. is governed by a large set of energy conserving scattering processes. This leads to diverging perturbative corrections at any order of perturbation theory. 
The bosonized fermionic interaction is quadratic in the Luttinger fields, while the band curvature transforms into a cubic nonlinearity $\propto\frac{1}{m}$. 

In the following, we will consider a nonequilibrium scenario in terms of an interaction quench. Initially, the system is supposed to be prepared in the ground state of the integrable Luttinger liquid theory at an interaction potential $g^i(x)$.
Due to the interaction quench, the interaction potential is suddenly switched at time $t=0$ from an initial to a final value
\eq{Eq4}{
g(x)=\left\{\begin{array}{c} g^i(x)\mbox{ for } t<0\\ g^f(x)\mbox{ for } t>0 \end{array}\right.
}
and both the quadratic Hamiltonian as well as the non-linearity are modified by this interaction change.
The eigenbasis of $H\sub{LL}$, which is expressed in terms of the physically more transparent phononic creation and annihilation operators $\creo{a}{q}, \anno{a}{q}$ according to the canonical Bogoliubov transformation 
\eq{Bog1}{
\theta_x &=& \theta_0 +\frac{i}{2}\int_q\left(\frac{2\pi}{|q|K}\right)^{1/2} e^{-iqx-\frac{|q|}{\Lambda}}\left(\creo{a}{q}-\anno{a}{-q}\right),\\
\phi_x&=&\phi_0-\frac{i}{2}\int_{q}\left(\frac{2\pi K}{|q|}\right)^{1/2}\mbox{sgn}(q)\ e^{-iqx-\frac{|q|}{\Lambda}}\left(\creo{a}{q}+\anno{a}{-q}\right)\label{Bog2},} is therefore obviously transformed by the quench. This transformation depends on the interaction via the Luttinger parameter $K$. 

The state of the system before the quench corresponds in general no longer to an equilibrium state after the quench, and the system will consequently undergo a nontrivial time evolution according to the new Hamiltonian. The occupations of bosonic modes after the quench can be computed via the above Bogoliubov transformation. Before the quench, the interacting system is in equilibrium at zero temperature, such that $G^K_{q,t=0}=\langle \{\anno{a}{q},\creo{a}{q}\}\rangle=1$ in the prequench basis. This yields the postquench occupations
\begin{eqnarray}
n_{t=0,q}&=&\langle \creo{a}{q}\anno{a}{q}\rangle_{t=0}=\frac{1}{2}\left[\frac{\lambda^2+1}{\lambda}n_{i,q}+\frac{\left(\lambda-1\right)^2}{2\lambda}\right],\nonumber\\
m_{t=0,q}&=&\langle\creo{a}{q}\creo{a}{-q}\rangle_{t=0}=\frac{1-\lambda^2}{4\lambda}\left(2n_{i,q}+1\right), \mbox{ with } \lambda=\frac{K\sub{f}}{K\sub{i}}.\ \ \ \ \ \ \ \ \ \label{Occ}
\end{eqnarray}
Here, $n_{i,q}$ is the initial occupation of the bosonic modes and $\lambda=\frac{K\sub{f}}{K\sub{i}}$ the ratio between the final $K\sub{f}=\sqrt{1+\frac{g\sub{f}}{\pi v\sub{F}}}$ and the initial $K\sub{i}=\sqrt{1+\frac{g\sub{i}}{\pi v\sub{F}}}$ Luttinger parameter. 
In this work, we focus on a zero temperature initial state, $n_{i,q}=0$ for all $q$. The phonon density after the quench $n_{t,q}> 0$ is always larger than the density before the quench, resulting in a nonzero excitation energy $\Delta E=\langle H\sub{f}\rangle-\langle H\sub{i}\rangle>0$ generated by the quench. Non-zero off-diagonal occupations $m_{t,q}\neq0$ indicate that the correlations are not diagonal in the post-quench quasi-particle basis and in order to relax to an equilibrium state, $m_{t,q}$ must decay to zero. In the present setting, we choose $m_{t,q}=e^{-2iu|q|t}\langle\creo{a}{q}\creo{a}{-q}\rangle_{t}$, such that the off-diagonal occupations remain always real, being either positive or negative, depending on the quench.

In the phonon basis, 
\eq{Eq5a}{
H\hspace{-1mm}=\hspace{-1.2mm}\int_q \hspace{-1.2mm} u|q|\creo{a}{q}\anno{a}{q}\hspace{-0.5mm}+\hspace{-1mm}\int_{q,k}\hspace{-3mm}\sqrt{|qk(k+q)|}\ v(k,q) \left(\creo{a}{q+k}\anno{a}{q}\anno{a}{k}+\mbox{h.c.}\right),}
with the vertex function $v(k,q)=v\left(\frac{q}{|q|},\frac{k}{|k|},\frac{k+q}{|k+q|}\right)$, which depends on the signs of the in- and outgoing momenta. In the interaction representation the phonon scattering Hamiltonian is
\eq{Eq5b}{
H\hspace{-1mm}_I(t)=\hspace{-1mm}\int_{q,k}\hspace{-3mm}\sqrt{|qk(k+q)|}\ v(k,q)\left(\creo{a}{q+k}\anno{a}{q}\anno{a}{k}e^{iut(|q+k|-|q|-|k|)}+\mbox{h.c.}\right).
}
Instead of solving the full problem, we aim at extracting the dominant contributions of the nonlinearity which are relevant for intermediate and large times and which drive the crossover towards thermalization.
In view of Eq.~\eqref{Eq5b}, off-resonant processes, for which $|q|+|k| \not= |k+q|$, will dephase and as a consequence become negligible for the intermediate and long-time evolution of the system~\cite{zwerger06}. Resonant processes on the other hand, here set by $|q|+|k| = |k+q|$, will at intermediate and long times become relevant in the renormalization group sense, as discussed in Ref.~\cite{Heyl2015nd}. 
\changed{The off-resonant processes can be eliminated perturbatively \cite{Heyl2015nd}, yielding subleading corrections for intermediate and large times, which we will neglect in the following. For the asymptotic thermalization process, these subleading corrections will yield non-universal corrections (i.e. observable in microscopic constants and prefactors). For instance, the presence of off-resonant scattering events will eventually lower the asymptotic temperature compared to a system with purely resonant scattering events.  The influence of off-resonant interactions on the decay rate of the bosonic and fermionic quasi-particles  has been investigated in Ref.~\cite{Proto14a}. The decay rate extracted from this computation is orders of magnitude lower than the rate due to purely resonant scattering processes.  Furthermore, it has a subleading scaling behavior $\sim Tq^4$ compared to  $\sim \sqrt{q^3T}$ for resonant scattering processes at small momenta $q$ \cite{zwerger06, andreev80}. Consequently, it is thus no influence on the leading order long time behavior. This allows us for the present purpose to restrict the phonon scattering to the resonant processes alone:}
\eq{Eq5}{
H\hspace{-1mm}=\hspace{-1.2mm}\int_q \hspace{-1.2mm} u|q|\creo{a}{q}\anno{a}{q}\hspace{-0.5mm}+\hspace{-1mm}v_0\int_{q,k}'\hspace{-3mm}\sqrt{|qk(k+q)|} \left(\creo{a}{q+k}\anno{a}{q}\anno{a}{k}+\mbox{h.c.}\right),}
where the integral $\int_{q,k}'$ is performed for momenta $|q+k|=|q|+|k|$ and $v_0=v(1,1)=\frac{3}{m}\sqrt{\frac{\pi}{K}}$ is the strength of the nonlinearity at resonance \cite{buchholdmethod,zwerger06}.

As we are interested in fermionic correlation functions, we switch from an operator based formalism to a field theoretical formulation on the Keldysh contour, which is explained in the appendix \ref{appendix2}, see also Ref.~\cite{buchholdmethod}. This allows us to treat both spatial and temporal forward time correlations on an equal footing. We will focus our analysis on the so-called fermionic lesser Green's function 
\eq{Green}{G^{<}_{t,x}=-i\langle\cre{\psi}{t,x}\anno{\psi}{t,0}\rangle} at equal forward times $t$ from which all fermionic equal time correlations can be deduced. Especially, in terms of a physical interpretation it is the Fourier transform of the fermionic momentum distribution 
\eq{Mom}{
n^{\mbox{\tiny F}}_{t,q}=i\int_x e^{iqx}G^<_{t,x}.
}
In the field theory representation, the bosonized fermionic lesser Green's function at equal times is
\eq{GF1}{
G^{<}_{\eta,t,x}=-i\langle\cre{\psi}{\eta,-,t,x}\anno{\psi}{\eta,+,t,0}\rangle=-i\Lambda\frac{e^{-i\eta k\sub{F}x}}{2\pi}e^{-\frac{i}{2}\mathcal{G}^<_{\eta,t,x}}.
}
Here, $\cre{\psi}{\nu},\ann{\psi}{\nu}$ label Grassmann fields with the index $\nu=(\eta,\gamma,t,x)$ representing right and left movers ($\eta=\pm$), the contour variables on the Keldysh plus and minus contour ($\gamma=\pm$), the forward time coordinate $t$ and the relative spatial distance $x$. The corresponding lesser  exponent $\mathcal{G}^<$ is defined as 
\eq{GF2}{
\mathcal{G}^<_{\eta, t,x}=2i\log\left\langle e^{i\left(\eta\phi_{+,t,0}-\theta_{+,t,0}-\eta\phi_{-,t,x}+\theta_{-,t,x}\right)}  \right\rangle.
}
The extra index $(\pm)$ of the Luttinger fields labels position on the plus-minus contour, see appendix. Combining Eq.~\eqref{GF2} and the Bogoliubov transformation above, one finds that $\mathcal{G}^{<}_{-\eta,t,x}=\mathcal{G}^{<}_{\eta,t,-x}$. The Green's function of the left movers is the spatially mirrored Green's function of the right movers, and it is sufficient to consider only the Green's function of the right movers 
\eq{GF3}{
G^{<}_{t,\eta x}\equiv G^{<}_{+,t,\eta x}=G^{<}_{\eta,t,x}
}
and equivalently for the exponent $\mathcal{G}^<$. According to the linked cluster theorem, the logarithm in Eq.~\eqref{GF2} is defined as the sum of all connected diagrams in an expansion of the exponent. As a consequence, it can be expressed to leading order in terms of the full Green's functions, with the next non-vanishing  correction being proportional to the equal-time one-particle irreducible four-point vertex, which is zero in the microscopic theory. Its effective correction remains negligibly small. In particular, the four-point vertex will only contribute to $\mathcal{O}[(um)^{-4}]$ which is two orders of magnitude smaller than the desired accuracy and its contribution can be safely neglected. The static one-particle irreducible four-point vertex represents a negligible correction for any equilibrium problem since it can only be generated via multiple concatenation of subleading three-point vertices. Especially it is not responsible for the modifications of the dynamic structure factor reported in Refs.~\cite{pustilnik07,imambekov09,lamacraft15}, since at zero temperature vertex corrections vanish exactly due to causality \cite{buchholdmethod,forsternelson76}. Consequently, the modifications of the dynamic structure factor happen entirely on the basis of the irreducible two-point vertex, i.e. the phonon self-energy. In the present case, the four-point vertex is exactly zero before the quench since this state corresponds to a zero temperature state as well as immediately after the quench, since a flat quasi-particle distribution in Eq.~\eqref{Occ} leads to a vanishing vertex correction.  
In terms of the Luttinger fields and apart from four-point vertex corrections, the exponent for the fermionic Green's function is
\eq{GF4}{
\mathcal{G}^<_{t,x}=\sum_{\alpha,\beta=\theta,\phi}\left(2\delta_{\alpha\beta}\hspace{-0.1cm}-\hspace{-0.1cm}1\right)\left[G^K_{\alpha\beta,t,0}\hspace{-0.1cm}
-\hspace{-0.1cm}G^K_{\alpha\beta,t,x}\hspace{-0.1cm}+\hspace{-0.1cm}G^A_{\alpha\beta,t,x}\hspace{-0.1cm}-\hspace{-0.1cm}
G^R_{\alpha\beta,t,x}\right]
,}
where $G^{R/A}_{\alpha\beta}$ is the retarded, advanced Green's function for $\alpha,\beta=\theta,\phi$ and $G^K_{\alpha\beta}$ is the corresponding Keldysh Green's function, i.e. $G^R_{\alpha\beta,t,x}=-i\langle \alpha_{q,x,t}\beta_{c,0,t}\rangle$.
Applying the Bogoliubov transformation to the phonon basis, the equal time exponent becomes
\begin{widetext}
\eq{GF5}{
\mathcal{G}^<_{t,x}=i\int_q\left[\mbox{$\frac{\pi e^{-\frac{|q|}{\Lambda}}}{|q|}$}(\cos(qx)-1)\left[\mbox{$\frac{K^2+1}{K}$}(2n_{t,q}+1)+2\mbox{$\frac{K^2-1}{K}$}\cos(2u|q|t)m_{t,q}\right]\right]+2\arctan(\Lambda x)+4i\int_q\left[\frac{\pi e^{-\frac{|q|}{\Lambda}}}{|q|}\sin(|q|x)\sin(2u|q|t)m_{t,q}\right].
}\end{widetext}
 Here, $n_{t,q}=\langle \creo{a}{t,q}\ann{a}{t,q}\rangle$ and $m_{t,q}=|\langle\ann{a}{t,-q}\ann{a}{t,q}\rangle|$ are the equal time normal and anomalous phonon densities, which evolve in time due to phonon scattering. The absence of the quasi-particle self-energy in this expression is caused by the equal time properties of the Green's function and underlines the fact that time-local, i.e. static, observables, even if explicitly forward-time dependent, are not modified by the frequency resolved fine structure of the self-energies once the time dependent distribution $n_{t,q}$ is known. In the remainder of this paper, we will analyze the time evolution of the exponent \eqref{GF5} after the interaction quench and its implications for the fermionic Green's function \eqref{GF1}.

Concerning the relevance of the interacting Luttinger model, before closing the section, we would like to mention that only recently pioneering experiments in ultra-cold gases both in and out of equilibrium explored the transient and prethermalization dynamics of systems~\cite{Hofferberth06,trupke13,schmiedmayer12,Langen14,nagerl13,meinert14,preiss15,hild14,cheneau12,agarwal14,schmiedmayernphys12,schmiedmayernjp13,bloch13,Guan2013} effectively described by a quadratic Luttinger model, the bosonic theory of the Hamiltonian in Eq.~\eqref{Eq2}. In particular, in Refs.~\cite{Hofferberth06,trupke13,schmiedmayer12,Langen14} prethermal states in the relative phase of a suddenly split condensate have been identified that have been stable on the experimentally accessible time scales.  For the latter experiments, the cubic nonlinearity studied in the present work constitutes the leading order correction to the quadratic theory in a gradient expansion. Therefore, the framework developed in the subsequent sections to describe the relaxation dynamics in the system, is of direct experimental relevance once the time scales are experimentally accessible to study the escape out of the prethermalization plateau. It is, however, important to note that the concrete experimental setup of the suddenly split condensate requires a further but straightforward extension of the considered model system to include two species of coupled bosonic fields. Moreover, let us emphasize that these experimental systems do not simulate the Luttinger liquid of interacting fermions -- our initial motivation -- but directly the effective bosonic low-energy theory. In this way, it might be possible to obtain experimental access to the dynamics of the bosonic occupation distributions, governed by the kinetic theory formulated below, via time-of-flight imaging.

\section{Summary of main results}
\label{sec:summaryofresults}

Before formulating and solving the kinetic theory for the interacting Luttinger liquid in detail, we briefly summarize the main results reported in this work. In the subsequent sections, we will then present the detailed calculations. Specifically, the known results on the purely integrable system are reformulated within the present framework in Sec.~\ref{sec:PT}. The kinetic equation, used to address the presence of the nonlinear phonon scattering, is derived in Sec.~\ref{sec:kinetic_equation}. This kinetic equation is then solved in Sec.~\ref{sec:thermalization_dynamics}.

It is the aim of this work to study the thermalization dynamics of the fermionic equal time Green's function \eqref{Green}, which is the Fourier transform of the fermionic momentum distribution \eqref{Mom} and contains the information on quadratic equal time fermion observables.
Without loss of generality, we focus on the distribution of the right-movers, i.e., $\eta=+$.  In the presence of phonon scattering, we determine the time-evolution of $G^<_{t,x}$ via a set of kinetic equations derived later in Sec.~\ref{sec:kinetic_equation}.

We find that $G^<_{t,x}$ features two distinct spatio-temporal crossover scales $x\sub{th}(t)$ and $x\sub{pt}(t)$, separating three regimes with distinct scaling behavior:
\renewcommand{\arraystretch}{1.5}
\begin{center}
\begin{tabular}{l l c }
	1. & prequench: $\quad$	&	$x\sub{pt}(t) \ll |x|$, \\
	2. & prethermal:	&	$x\sub{th}(t) \ll |x| \ll x\sub{pt}(t)$, \\
	3. & thermal:		&	$|x| \ll x\sub{th}(t)$. \\
\end{tabular}
\end{center}
\renewcommand{\arraystretch}{1}
We find for the associated crossover scales $x_\mathrm{pt}(t)$ and $x_\mathrm{th}(t)$:
\eq{eq:crossoverScales}{
x_\mathrm{pt}(t) = 2ut, \qquad \qquad x_\mathrm{th}(t) =\frac{x_{\lambda}}{\Lambda} \left( v_0 \Lambda^2 t \right)^{\alpha_\lambda}.
}
The first crossover at $x\sub{pt}(t)$ determines the light cone~\cite{cardycalabrese06} set by the sound velocity $u$ of the phononic elementary excitations and is known from the non-interacting Luttinger model. Two space points a distance $x \gg x\sub{pt} (t)$ apart from each other have not been able to exchange information after the quench due to causality. Therefore, the properties at such distances are solely given by the initial condition before the quench such that we term this regime ``prequench''. \changed{For distances $x<x_{\text{pt}}$ quasi-particles are starting to form, marking the onset of prethermalization.} 
The second crossover takes place at $x=x\sub{th}(t)$ setting the scale for the onset of thermalization due to quasi-particle scattering. The exponent $\alpha_{\lambda}$ with $0<\alpha_{\lambda}<1$, as well as the dimensionless length $x_{\lambda}$, depends on the quench parameter $\lambda$ only and can be determined numerically. The upper bound of $\alpha_{\lambda}$ is guaranteed by the subleading nature of the vertex, which forbids ballistic spreading in the thermal region.
\changed{The treatment of quasi-particle scattering in terms of a kinetic equation approach is only valid on distances, for which a well-defined prethermal plateau has been established. Given this, we estimate the kinetic equation approach to be valid on distances
\eq{dist}{
x<x_c(t)=x_{\text{th}}(t)\exp\left(-\tfrac{K^2+1}{|K^2-1|}\sqrt{\tfrac{3n_{\lambda}}{|m_{\lambda}|}}\right)
}
and in the scattering-less region $x>x_{\text{th}}$. In the intermediate regime $x_c(t)<x<x_{\text{th}}$, quasi-particle scattering is as fast as the formation of quasi-particles, such that both effects have no distinguishable time scale. While the results obtained from our approach might not be reliable in this region, $x_{\text{th}}(t)$ remains the crossover scale below which the non-linearity becomes non-negligible.}
The fact that $x\sub{th}(t)$ has an explicit dependence on the Luttinger cutoff $\Lambda$ ($\alpha_{\lambda}>1/2$ generally) is not surprising. The non-linearity in the Luttinger model introduces a microscopic energy scale $v_0\Lambda^2$ which represents the characteristic time scale of the dynamics induced by the non-linearity, i.e. in the present case the thermalization dynamics beyond the quadratic theory. Additionally, the non-linearity breaks the scale invariance of the quadratic model, which is responsible for the fact that all microscopic scales can be eliminated from macroscopic observables in that case. In the absence of scale invariance, however, the microscopic length scale $\Lambda$ will appear in certain observables, expressing that their explicit value depends on model specific details. 

As we show in our detailed analysis below, we find that this separation into three spatio-temporal regimes -- prequench, prethermal, and thermal -- reflects itself in a remarkable factorization property of the Green's function
\eq{eq:factorization2}{
G^<_{t,x}=G^<_{0,x}Z\sub{pt}(s\sub{pt})Z\sub{th}(s\sub{th}),}
which holds everywhere except in the vicinity of the crossover scales $x_\mathrm{th}(t)$ and $x_\mathrm{pt}(t)$.  Here, we have introduced the following short-hand notations:
\begin{equation}
s\sub{pt}=\left\{\begin{array}{cl} x &\mbox{ for } x<x_\mathrm{pt}(t)\\ 2ut& \mbox{ for } x>x_\mathrm{pt}(t)\end{array}\right. , \quad s\sub{th}=\left\{\begin{array}{cc}x&\mbox{ for } x<x\sub{th}(t)\\
x\sub{th}(t)&\mbox{ for } x>x\sub{th}(t)\end{array}\right. .
\end{equation}
While the factorization into $G^<_{0,x}$ and $Z\sub{pt}$ has been already known for the exact solution of the integrable model~\cite{cazalilla06}, here, we show that the influence of the nonlinearity can be captured by a further factor in terms of $Z\sub{th}$. The thermal contribution $Z\sub{th}(s\sub{th})$ exhibits interesting spatio-temporal dynamics in particular in the long-time regime $ut \gg x\sub{th}(t)$. It is defined as
\begin{equation}
  Z_\mathrm{th} (s_\mathrm{th}) = \exp\left(-\frac{K^2+1}{K} \frac{\pi \tilde T_t|s_\mathrm{th}|}{u}\right)
\end{equation}
and features two different spatio-temporal regimes.

\emph{(i) thermalized regime:}
Deep in the thermalized region $|x|\ll x\sub{th}(t)$ where $s\sub{th}=x$, $Z\sub{th}=\exp(-|x|/\xi_{\tilde{T}_t})$ exhibits the conventional exponential decay with distance that the system experiences in thermal states with an associated thermal length
\begin{equation}\label{tlength}
	\xi_{\tilde T_t} = \frac{K}{1+K^2} \frac{u}{\pi \tilde T_t}.
\end{equation}
The effective temperature $\tilde T_t$, however, entering this equation remains a dynamical quantity with
\begin{equation}
  \tilde T_t=T+u \Lambda \Delta_\lambda(v_0 \Lambda^2 t)^{-\mu},
\end{equation}
approaching the temperature $T$ of the final thermal ensemble algebraically slowly. We find that the numerical simulations of the kinetic equation are consistent with an analytical estimate for the exponent $\mu=2/3$. Thus, the system in this spatial region appears to be hotter than in the final asymptotic thermal state. The associated excess energy stored at short distances has to be transported to larger distances which, however, is an algebraically slow process since this energy transport in the presence of detailed balance is carried out by dynamical slow modes, emerging as a consequence of exact conservation laws \cite{Lux13}.

\emph{(ii) prethermal and prequench regime:} Within the prethermal and prequench region $x\sub{th}(t) \ll x$, the amplitude $Z\sub{th}(s\sub{th})=Z\sub{th}(x\sub{th}(t))$ approaches a space-independent but time-dependent constant quantifying the temporal decay of the prethermal correlations:
\begin{equation}
  Z\sub{th}(x\sub{th}(t)) = \exp[-x\sub{th}(t)/\xi_{\tilde T_t}].
\end{equation}
Because $x\sub{th}(t) \propto (v_0 \Lambda^2 t)^{\alpha_\lambda}$, we have, remarkably, that this amplitude decays in stretched exponential form. This decay is sub-exponential and thus inherently nonperturbative in nature, highlighting the capabilities of our present approach.

\section{Dynamics in the absence of phonon scattering}
In order to systematically understand the effect of phonon scattering on the relaxation dynamics after the interaction quench, we first determine the dynamics of the exponent $\mathcal{G}^<_{t,x}$ in the absence of scattering, i.e. for $\frac{1}{m},v_0\rightarrow 0$. This quench scenario has been extensively discussed in \cite{Karrasch12,Kennes13,cazalilla06,Iucci09,Iucci10}, and we will only briefly list the known results in the present formalism in order to make contact to the relaxation dynamics in the presence of phonon scattering, which are discussed subsequently.
\subsection{Ground state properties}
For a system in the ground state, $n_{t,q}=m_{t,q}=0$ and the exponent evaluates to
\eq{GF6}{
\mathcal{G}^<_{t,x}=-i\frac{K^2+1}{2K}\log(1+\Lambda^2x^2)+2\arctan(\Lambda x),}
which leads to a time-independent fermionic Green's function
\eq{GF7}{
G^<_{t,x}=-\frac{i\Lambda}{2\pi}e^{-ik\sub{F}x-i\arctan(\Lambda x)}\sqrt{1+\Lambda^2x^2}^{-\frac{K^2+1}{2K}}
,}
well known from the literature \cite{haldane81,giamarchi04}. It features an algebraic decay in space $\sim x^{-\frac{K^2+1}{2K}}$ and a power law singularity of the fermionic momentum distribution close to the Fermi momentum $n^{\mbox{\tiny F}}_q\sim|q-k\sub{F}|^{-\frac{(K-1)^2}{2K}}$ \cite{giamarchi04}. 
\subsection{Quench from the ground state}\label{sec:PT}
Initializing the fermions in the ground state and performing an interaction quench leads to constant non-zero phonon densities in the post-quench basis, according to Eq.~\eqref{Occ}. In the absence of scattering, the phonon densities are constants of motion and remain time independent, $n_{t,q}=n_{0,0}\equiv n$ and $m_{t,q}=m_{0,0}\equiv m$. In this situation, only dephasing of the off-diagonal Green's functions induces relaxation and the exponent is
\begin{widetext}\begin{eqnarray}
\label{GF8}
\mathcal{G}^<_{t,x}&=&2\arctan(\Lambda x)-i\mbox{$\frac{K^2+1}{2K}$}(2n+1)\log(1+\Lambda^2x^2)+im\log\left(\mbox{$\frac{1+\Lambda^2(x-2ut)^2}{1+\Lambda^2(x+2ut)^2}$}\right)-i\mbox{$\frac{K^2-1}{2K}$}m\left[\log\left(\mbox{$\frac{1+\Lambda^2(x-2ut)^2}{1+4u^2t^2\Lambda^2}$}\right)+\log\left(\mbox{$\frac{1+\Lambda^2(x+2ut)^2}{1+4u^2t^2\Lambda^2}$}\right)\right]\nonumber\\
&=&\mathcal{G}^<_{0,x}+im\log\left(\mbox{$\frac{1+\Lambda^2(x-2ut)^2}{1+\Lambda^2(x+2ut)^2}$}\right)-i\mbox{$\frac{K^2-1}{2K}$}m\log\left[\mbox{$\frac{(1+\Lambda^2(x+2ut)^2)(1+\Lambda^2(x-2ut)^2)}{(1+4u^2t^2\Lambda^2)^2(1+x^2\Lambda^2)^2}$}\right].\end{eqnarray}\end{widetext}
Here, $\mathcal{G}^<_{0,x}$ is the exponent corresponding to the prequench state, i.e. the ground state of interacting fermions with the prequench Luttinger parameter $K_i$. Consequently the fermion Green's function \eqref{GF1} factorizes
\eq{GF9}{
G^<_{t,x}=G^<_{0,x}\tilde{Z}\sub{pt}(x,t).
}
The factor $\tilde{Z}\sub{pt}$ is defined by Eqs.~\eqref{GF8} and \eqref{GF1} and describes the time-dependent modification of the initial zero temperature Green's function due to the quench. In view of the following discussion it is useful to investigate this factor on distances away from the light cone $x=2ut$. For distances $|x|\ll 2ut$, the temporal factors in Eq.~\eqref{GF8} cancel each other and $\tilde{Z}\sub{pt}(t,x)\overset{|x|\ll2ut}{\rightarrow} Z\sub{pt}(x)$ looses its time dependence. On the other hand, for distances $|x|\gg2ut$, the spatial dependence drops out and $\tilde{Z}\sub{pt}(t,x)\overset{|x|\gg2ut}{\rightarrow} Z\sub{pt}(2ut)$. This defines the prethermal amplitude
\eq{GF10}{
Z\sub{pt}(s)=\left(\sqrt{1+\Lambda^2s^2}\right)^{\frac{K^2-1}{2K}\frac{1-\lambda^2}{4\lambda}}.
}
The process associated with the crossover of $Z\sub{pt}(s)$ from a temporal to a spatial dependence as a function of time is the formation of quasi-particles corresponding to the post-quench Hamiltonian. This is the typical prethermalization scenario in the absence of quasi-particle scattering. For short times, the properties of the system are dominated by the initial state of the system, and the fermion Green's function is only modified by a global amplitude but has the same spatial scaling behavior as for the initial state. The effect of the quadratic Hamiltonian in the time evolution is the dephasing of all terms, which are not diagonal in the basis of the post-quench quasi-particles, leading to a diagonal ensemble in the quasi-particles with a non-equilibrium phonon density. This non-equilibrium distribution of phonons induces a scaling behavior of the fermion Green's function in real space, which is different from the zero and finite temperature cases.

In the absence of phonon scattering, the diagonal phonon densities $n_{t,q}$ are constants of motion and do not relax, the density matrix $\rho$ therefore does not approach a Gibbs state but is rather described in the asymptotic limit $t\rightarrow\infty$ by a generalized Gibbs ensemble (GGE), which respects the constants of motion and maximizes the entropy. It is given by
\eq{GF11}{
\rho\sub{GGE}=Z\sub{GGE}^{-1}e^{-\int_q \nu_q \hat{n}_q},}
where the Lagrange parameters $\nu_q=2\log\left(\frac{\lambda+1}{|\lambda-1|}\right)$ depend on the quench parameter and $Z\sub{GGE}$ is the normalization factor.

The fermion Green's function for the two different regimes is then
\eq{GF12}{
G^<_{t,x}=G^<_{0,x}\times\left\{\begin{array}{ll} Z\sub{pt}(2ut) &\mbox{ for } |x|\gg 2ut\\
Z\sub{pt}(x)& \mbox{ for } |x|\ll 2ut\end{array}\right. ,
}
with the non-equilibrium scaling behavior 
\eq{GF13}{G^<_{t,x}\overset{t\rightarrow\infty}{\sim} |x|^{-\gamma\sub{Eq}\gamma\sub{GGE}},
}
where $\gamma\sub{Eq}=\frac{K^2+1}{2K}$ is the equilibrium exponent and $\gamma\sub{GGE}=\frac{\lambda^2+1}{2\lambda}=2n+1$ (see, Eq.~\eqref{Occ}) is the non-equilibrium correction resulting from a non-thermal quasi-particle distribution.
\section{Phonon scattering and the kinetic equation}
\label{sec:kinetic_equation}

In the previous sections, we have demonstrated that the forward time evolution of the fermionic equal-time Green's function can be determined solely from the momentum dependent excitation distributions $n_{t,q}, m_{t,q}$. All quadratic, equal-time observables on the other hand can be computed from the fermionic equal-time Green's function via a unitary transformation, such that the knowledge of $n_{t,q}$ and $m_{t,q}$ gives access to the forward time evolution of all the frequency independent quadratic fermion observables. Therefore the time-evolution of this specific set of observables can be captured by the time evolution of the frequency independent and well-defined quantities $n_{t,q}, m_{t,q}$, which does not necessitate the frequency resolved fine structure in the fermionic spectrum. 
In order to determine the time-evolution of the phonon densities, we derive kinetic equations for the excitation distribution function \cite{kamenevbook} in the limit of well defined excitations, closely following the steps in Ref.~\cite{buchholdmethod} and briefly discussing the approximations. 

Before we start with the explicit derivation, we review very briefly the known results for nonlinear Luttinger liquids (c.f. \cite{imambekov12}) and place the present approach into this context. At zero temperature and without band curvature, long wavelength physics of the interacting fermion model can be exactly mapped to the quadratic Luttinger model and therefore has well-defined, sharp phononic excitations, expressed by a spectral function of the phonons $\mathcal{A}_{q,\omega}=i(G^R_{q,\omega}-G^A_{q,\omega})=2\pi \delta(\omega-u|q|)$. In the presence of band curvature, however, the phonons themselves interact via a resonant three-point scattering vertex, which leads to a broadening of the spectral function around the mass-shell $\omega=u|q|$. This broadening can be described in terms of two excitations branches at frequencies $\omega=\epsilon^{\pm}_q$, where $\epsilon^-_q<u|q|$ labels a solitonic branch and $\epsilon^+_q>u|q|$ labels a phononic branch (such that $|\epsilon^{+}_q-\epsilon^-_q|/q\rightarrow 0$ for $q\rightarrow0$)\cite{imambekov12,lamacraft15}. The spectral weight of the excitations in the nonlinear Luttinger liquid is distributed continuously between these two branches. Whereas the solitonic branch represents an exact boundary (i.e. no spectral weight is located at frequencies $\omega<\epsilon_q^-$), featuring a power law singularity for frequencies above $\epsilon^-_q$, the phononic branch represents an algebraically sharp boundary (i.e. the spectral weight for frequencies $\omega>\epsilon_q^+$ is strongly algebraically suppressed), featuring a power law singularity from both sides \cite{imambekov12}. While the power-law singularities at the edges of the spectral weight obviously cannot be explained by a frequency independent self-energy, the characteristic width of the spectral weight $\delta\omega_q=\epsilon^+_q-\epsilon^-_q=\frac{q^2}{m^*}$ can be captured by an imaginary part of the on-shell value of the self-energy $\Sigma^R_{q,\omega=u|q|}$, which determines the renormalized mass $m^*$ \cite{zwerger06,aristov,lamacraft15,imambekov12}. These results hold for the zero temperature limit of the problem. At finite temperature $T>0$, however, a self-consistent Born-approximation for the on-shell self-energy predicts a scaling of the spectral weight $\delta\omega_q\sim \sqrt{|q|^3T}$ \cite{lamacraft13,gangardt13}, which has been also observed in numerical simulations of interacting one-dimensional bosons \cite{lamacraft13}. For $\delta\omega_q\ll u|q|$, i.e. the width of the spectral weight of the excitations being much smaller than the average excitation energy, the spectral weight is still sharply concentrated at the mass-shell and one can still think (physically) of well defined excitations although the fine structure of the spectral weight is very different from what one is used to for weakly interacting quasi-particles. As a consequence, it is possible to derive a kinetic equation for the excitation densities in this regime, applying the common quasi-particle and local time approximations, and we will implement this approach below. It neglects the specific structure of the spectral weight of nonlinear Luttinger liquids, which is valid for "static" variables in the quasi-particle limit $\delta\omega_q\ll u|q|$.
We begin by introducing the interaction picture for the Heisenberg operators 
\eq{Eq21}{
\cre{a}{t,q}\rightarrow \cre{a}{t,q}e^{-iu|q|t},}
which leaves the Hamiltonian \eqref{Eq5} unmodified but shifts the spectral weight of diagonal modes to zero frequency and eliminates the phase $\sim e^{i2u|q|t}$ of off-diagonal correlation functions \cite{buchholdmethod}. 
 The Green's functions in the interaction representation are labeled with a tilde. The Keldysh Green's function in Nambu space is
\eq{Eq22}{
i\tilde{G}^K_{t,q,\delta}=\left(\hspace{-0.15cm}\begin{array}{ll}\langle\{\ann{a}{t+\frac{\delta}{2},q},\cre{a}{t-\frac{\delta}{2},q}\}\rangle & \langle\{\ann{a}{t+\frac{\delta}{2},q},\ann{a}{t-\frac{\delta}{2},-q}\}\rangle\\
\langle\{\cre{a}{t+\frac{\delta}{2},-q},\cre{a}{t-\frac{\delta}{2},q}\}\rangle&\langle\{\cre{a}{t+\frac{\delta}{2},q},\ann{a}{t-\frac{\delta}{2},q}\}\rangle
\end{array}\hspace{-0.15cm}\right),\ \ \ 
}
where  $\{\cdot,\cdot\}$ is the anti-commutator and we introduced an additional relative time shift $\delta$ associated with spectral properties of the system. The retarded Green's function is
\eq{Eq23}{
i\tilde{G}^R_{t,q,\delta}&=&\theta(\delta)\left(\hspace{-0.15cm}\begin{array}{ll}\langle[\ann{a}{t+\frac{\delta}{2},q},\cre{a}{t-\frac{\delta}{2},q}]\rangle & \langle[\ann{a}{t+\frac{\delta}{2},q},\ann{a}{t-\frac{\delta}{2},-q}]\rangle\\
\langle[\cre{a}{t+\frac{\delta}{2},-q},\cre{a}{t-\frac{\delta}{2},q}]\rangle&\langle[\cre{a}{t+\frac{\delta}{2},q},\ann{a}{t-\frac{\delta}{2},q}]\rangle
\end{array}\hspace{-0.15cm}\right)\nonumber\\
&=&\theta(\delta)\left(\hspace{-0.15cm}\begin{array}{cc}\langle[\ann{a}{t+\frac{\delta}{2},q},\cre{a}{t-\frac{\delta}{2},q}]\rangle & 0\\
0&\langle[\cre{a}{t+\frac{\delta}{2},q},\ann{a}{t-\frac{\delta}{2},q}]\rangle
\end{array}\hspace{-0.15cm}\right).}
The off-diagonal retarded and advanced Green's functions are exactly zero. This is a consequence of the Hamiltonian, which does not introduce a coupling between the modes $q$ and $-q$, such that the commutator $[\ann{a}{t+\frac{\delta}{2},q},\ann{a}{t-\frac{\delta}{2},-q}]=0$ for all times $t,\delta$. The anti-hermitian Keldysh Green's function is parametrized according to \cite{kamenevbook,buchholdmethod}
\eq{Eq24}{
\tilde{G}^K_{t,q,\delta}=\left(\tilde{G}^R\circ\sigma_z\circ F-F\circ\sigma_z\circ\tilde{G}^A\right)_{t,q,\delta}}
in terms of the time-dependent, hermitian quasi-particle distribution function $F$ and the Pauli matrix $\sigma_z$, the latter preserving the symplectic structure of bosonic Nambu space. The $\circ$ represents matrix multiplication with respect to momentum space and convolution with respect to time. Switching to Wigner coordinates by Fourier transforming the Keldysh Green's function with respect to relative time
\eq{Eq25}{
\tilde{G}^K_{t,q,\omega}=\int_{\delta}\tilde{G}^K_{t,q,\delta}\ \ e^{i\omega\delta}
}
and applying the Wigner approximation, which, due to the RG-irrelevant interactions, is justified in the same regime for which the Luttinger description is applicable \cite{buchholdmethod,Heating}, we find
\eq{Eq26}{
\tilde{G}^K_{t,q,\omega}=\tilde{G}^R_{t,q,\omega}\sigma_zF_{t,q,\omega}-F_{t,q,\omega}\sigma_z\tilde{G}^A_{t,q,\omega},} 
which is diagonal in momentum and frequency space. Inverting Eq.~\eqref{Eq26} by multiplying it with $\left(\tilde{G}^R\right)^{-1}$ from the left and $\left(\tilde{G}^A\right)^{-1}$ from the right, yields the kinetic equation for the distribution function
\eq{Eq27}{
i\partial_tF_{t,q,\omega}\hspace{-0.1cm}=\hspace{-0.1cm}\sigma_z\Sigma^R_{t,q,\omega}F_{t,q,\omega}\hspace{-0.1cm}-\hspace{-0.1cm}F_{t,q,\omega}\Sigma^A_{t,q,\omega}\sigma_z\hspace{-0.1cm}-\hspace{-0.1cm}\sigma_z\Sigma^K_{t,q,\omega}\sigma_z.\ \ \ \ \ \ 
}
The retarded, advanced self-energies $\Sigma^{R/A}_{t,q,\omega}$ are diagonal in Nambu space, while the Keldysh self-energy $\Sigma^K_{t,q,\omega}$ consists of non-vanishing diagonal and off-diagonal entries due to the initial off-diagonal occupations $m_{0,q}\neq0$.

The kinetic equation for the phonon occupations is obtained by multiplying Eq.~\eqref{Eq27} on both sides with the spectral function $\tilde{\mathcal{A}}_{t,q,\omega}=i\left(\tilde{G}^R_{t,q,\omega}-\tilde{G}^A_{t,q,\omega}\right)$ and integrating over frequency space. For interacting Luttinger Liquids, the spectral function $\tilde{\mathcal{A}}_{t,q,\omega}$ is very narrowly peaked at the mass shell and the kinetic equation is essentially locked onto $\omega=0$ in this way (in the interaction picture, the mass shell is at $\omega=0$). As a consequence, one finds kinetic equations for the diagonal densities
\eq{Eq28}{
\partial_tn_{t,q}=-\sigma^R_{t,q}(2n_{t,q}+1)+\sigma^K_{t,q}
}
and the off-diagonal densities
\eq{Eq29}{
\partial_tm_{t,q}=-2\sigma^R_{t,q}m_{t,q}-\Gamma^K_{t,q}.
}
They can be expressed in terms of the imaginary part of the retarded on-shell self-energy \eq{Nun1}{\sigma^R_{t,q}=\frac{1}{2}\int_{\omega}\tilde{\mathcal{A}}_{t,q,\omega}\left(\Sigma^R_{t,q,\omega}-\Sigma^A_{t,q,\omega}\right)\approx\frac{1}{2}\left(\Sigma^R_{t,q,\omega=0}-\Sigma^A_{t,q,\omega=0}\right)} and the Keldysh on-shell self-energies \eq{Nun2}{\sigma^K_{t,q}=\frac{i}{2}\int_{\omega}\tilde{\mathcal{A}}_{t,q,\omega}\left(\Sigma^K_{t,q,\omega}\right)_{11}\approx\frac{i}{2}\left(\Sigma^K_{t,q,\omega=0}\right)_{11}} and \eq{Nun3}{\Gamma^K_{t,q}=\frac{i}{2}\int_\omega\mathcal{A}_{t,q,\omega}\left(\Sigma^K_{t,q,\omega}\right)_{12}\approx\frac{i}{2}\left(\Sigma^K_{t,q,\omega=0}\right)_{12}.} The Keldysh self-energy is always anti-hermitian and therefore purely imaginary in frequency and momentum space, such that Eqs.~\eqref{Eq28}, \eqref{Eq29} are real. Since the criterion $|\epsilon^+_q-\epsilon^-_q|\ll u|q|$ is equivalent to $\sigma^R_{t,q}\ll u|q|$ at zero and finite temperature equilibrium, we also apply the latter criterion for the present out-of-equilibrium situation in order to estimate the validity of our approach.

\begin{figure}
\centering
  \includegraphics[width=1\linewidth]{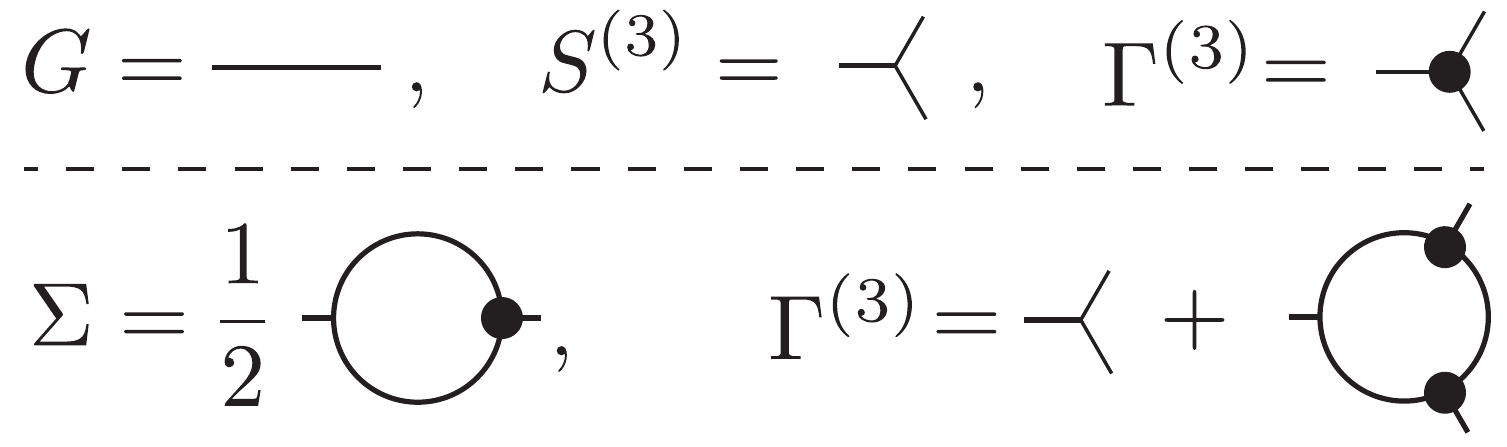}
  \caption{Diagrammatic illustration of the Dyson-Schwinger equations up to cubic order. Here, $G$ represents the full Green's function, $S^{(3)}$ the bare three-body vertex and $\Gamma^{(3)}$ the full three-body vertex. For convenience, this displays only the topology of the diagrams, which has not been extended to Keldysh space.}
  \label{fig:Diag}
\end{figure}

The phonon scattering terms in Eq.~\eqref{Eq5} are resonant, i.e. they describe scattering between a continuum of energetically degenerate states, and as a consequence, perturbation theory diverges. In order to determine the self-energies $\sigma^R_{t,q}, \sigma^K_{t,q}, \Gamma^K_{t,q}$, we apply non-perturbative Dyson-Schwinger equations, which are truncated at cubic order. This takes into account renormalization effects of the cubic vertex and yields non-perturbative self-energies. The topology of the corresponding diagrams is shown in Fig.~\ref{fig:Diag}.
 If we neglect the cubic vertex correction, the Dyson-Schwinger equations reduce to the self-consistent Born approximation \cite{buchholdmethod}. For an initial state with constant phonon density, as it is the case for the present setup, the vertex correction has been shown to be exactly zero \cite{buchholdmethod,forsternelson76}, however it obtains a non-zero value in the time-evolution of the system. The kinetic equations \eqref{Eq28}, \eqref{Eq29} are solved iteratively, starting at a certain time $t$, the self-energies and vertex correction are computed as functions of the distributions $n_{t,q}, m_{t,q}$. Subsequently $\partial_tn_{t,q}, \partial_tm_{t,q}$ are determined,  and used in turn to compute the distributions $n_{t+\Delta,q}, m_{t+\Delta,q}$ for an infinitesimally later time. This procedure is repeated in order to determine the time-evolution of the phonon densities and self-energies. A more detailed, technical derivation of the iterative solution for the kinetic equation, self-energies and vertex correction can be found in \cite{buchholdmethod}.

\section{Thermalization and Prethermalization Dynamics}
\label{sec:thermalization_dynamics}

As one can see from the kinetic equations in Eq.~\eqref{Eq28} and Eq.~\eqref{Eq29}, the diagonal and off-diagonal phonon densities are no longer constants of motion in the presence of phonon scattering and energy is redistributed between the different momentum modes. On a general level, when the system thermalizes, as we will show below, the steady state of the dynamics in the presence of a cubic scattering as in Eq.~\eqref{Eq5}, is solely determined by the associated temperature $T$ and independent of any further details of the initial nonequilibrium state. Specifically, the diagonal modes acquire a Bose-Einstein distribution $n_{\infty,q} = n_{t\rightarrow\infty,q}=\left(e^{ u|q|/T}-1\right)^{-1}$ whereas the off-diagonal distributions $m_{q}=0$ have to vanish.

Importantly, in the resonant approximation, the final temperature $T$ ($k\sub{B}=1$ in the following) can be computed directly from the initial state as will be shown now. In a closed system, the total energy is conserved. Moreover, the conservation of the kinetic energy is an additional exact feature of the derived kinetic equation. As a consequence, also the interaction energy itself is individually conserved. The latter is not an artifact of the kinetic equation but a feature of the resonant nature of the interactions, which, by definition of resonance, commute with the quadratic part of the Hamiltonian \eqref{Eq5} already on an operator level. 
This implies that the relaxation dynamics due to the interactions takes place in closed subsets of degenerate eigenstates of the quadratic Hamiltonian, which would in the absence of phonon scattering only acquire a global phase and were not able to thermalize. 
Consequently, the kinetic energy of the initial ($e_0$) and final state ($e_f$) have to be equal, which yields:
\eq{Eq20}{
e_{0}=un_{\lambda}\Lambda^2=\int_q \hspace{-0.1cm}u|q| n_{0,q}\overset{!}{=}\int_q u |q|n_{\infty,q}=\frac{T^2_{\lambda}\pi^2}{3u}=e_{f}.\ \ \ 
}
Here, $n_{0,q}$ is the initial momentum distribution, see Eq.~\eqref{Occ}, and $n_{\infty,q} = \left(e^{\beta u|q|}-1\right)^{-1}$ is the final, thermal distribution. This gives:
\eq{temperature}{T_{\lambda}=\frac{u\Lambda}{\pi}\sqrt{3n_{\lambda}},} which depends on the details of the quench only through the quench parameter $\lambda$ such that we denote the temperature via $T\sub{$\lambda$}$ in the following. Importantly, this temperature yields a criterion for the applicability of the Luttinger theory for the present quench scenario, since Luttinger theory is only well-defined for temperatures lower than the cutoff $T_{\lambda}<u\Lambda$. Evaluating this inequality results in a bound for the quench parameter $\lambda$, i.e. for $\frac{1}{15}\le \lambda\le15$, the quench can be described in the framework of Luttinger theory.

In the remainder of this section, we will discuss the time-evolution of the phonon densities according to the kinetic equation and derive the form of the Green's function in Eq.~\eqref{eq:factorization2}.
\subsection{Phonon densities}
The time evolution of the phonon densities is determined by the kinetic equations \eqref{Eq28} and \eqref{Eq29}. In order to make the time evolution of the phonon densities dimensionless, we rescale the self-energy according to $\tilde{\sigma}^{R,K}=\frac{\sigma^{R,K}}{v_0\Lambda^2}$, the momentum $\tilde{q}=\frac{q}{\Lambda}$ and time $\tau=v_0\Lambda^2t$. In these units, the time evolved phonon densities depend only on the initial state and are independent of the microscopic details of $v_0$ and $\Lambda$ \cite{buchholdmethod}, i.e. in the present setting the time-evolution of the phonon density is completely determined by the quench parameter $\lambda$, which characterizes the initial state. Additionally, as a consequence of Eq.~\eqref{Occ}, the dynamics remains invariant under $\lambda\rightarrow1/\lambda$ and $m_{\tau,q}\rightarrow-m_{\tau,q}$ and we therefore consider only the case $\lambda>1$.

\begin{figure*}
  \includegraphics[width=1\linewidth]{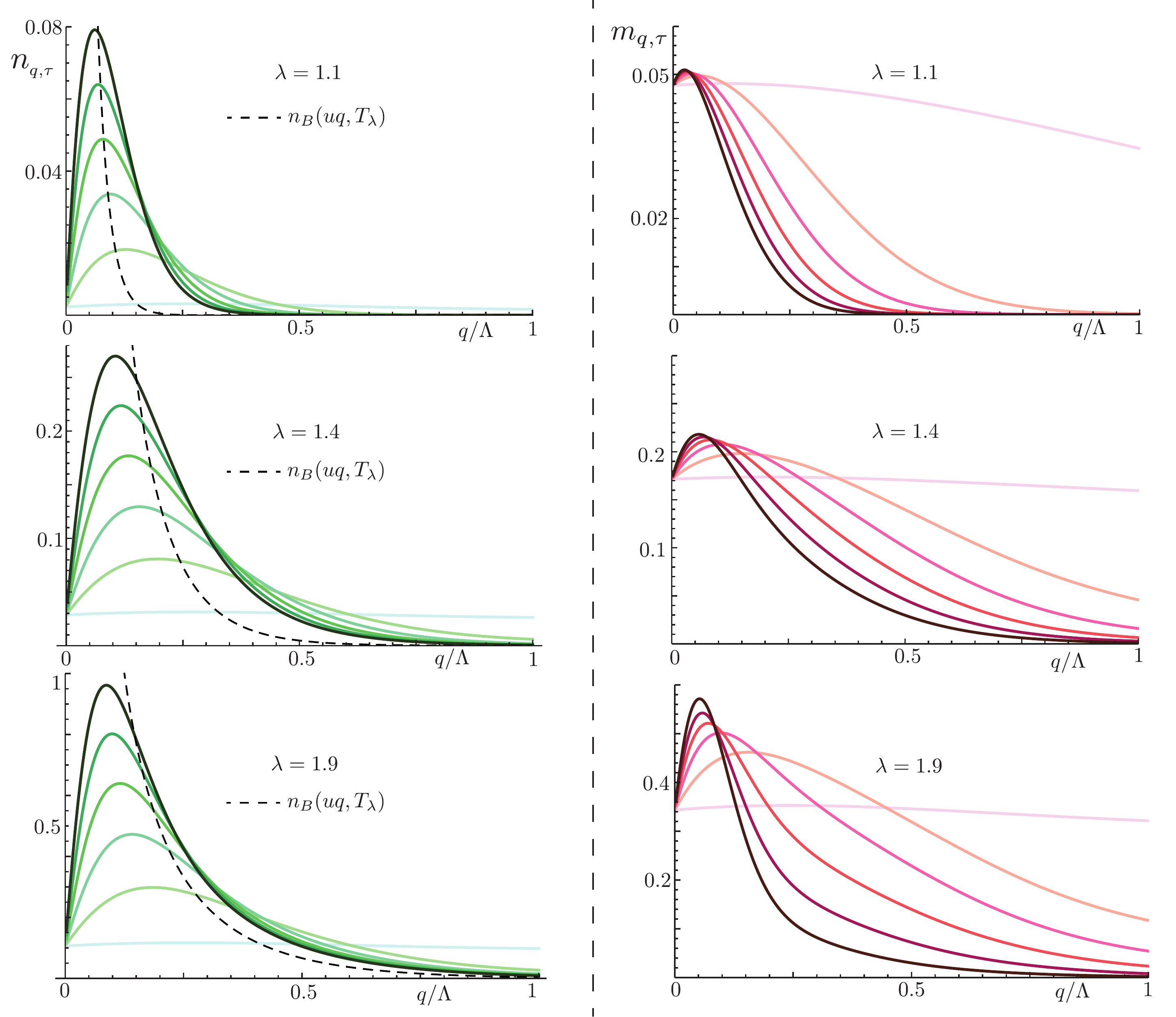}
  \caption{Simulation of the time-evolution of the diagonal phonon density $n_{\tau,q}$ (left column) and off-diagonal density $m_{\tau,q}$ (right column) for different quench parameters $\lambda$. In each row, the individual lines correspond to different times $\tau=(0,1,2,3,4,5)$. 
Left column: The total phonon density increases in time (from light to dark green) and the dotted lines represent the corresponding asymptotic density in the limit $\tau\rightarrow\infty$, which is a Bose distribution with the quench dependent temperature $T_{\lambda}=(0.035, 0.124, 0.24)u\Lambda$ (from the top to the bottom row). The distribution function is separated into two regimes according to Eq.~\eqref{Eq37}, with a linear increase in momentum for small momenta and a corresponding thermal distribution for larger momenta. The crossover momentum separating the two regimes is marked with a dot.
Right column: The off-diagonal phonon density is decreasing in time (from light to dark red), displaying two distinct momentum regimes: For momenta larger than the crossover, $q>q\sub{th}$, the off-diagonal occupation decreases exponentially in momentum, while it remains close to its initial value $m_{0,q}=m_{\lambda}$ for momenta smaller than the crossover. While any momentum mode $n_{\tau,q>0}$ will thermalize at a finite time $\tau<\infty$, the zero momentum mode remains pinned to its initial value $n_{\tau,q=0}=n_{\tau=0,q=0}$. The latter is not an artifact of the approximation but a consequence of exact fermionic particle number conservation, as outlined in the main text.
}
  \label{fig:OccFull}
\end{figure*}

The time evolution of the phonon densities for three different quench parameters $\lambda$ is shown in Fig.~\ref{fig:OccFull}. It features two characteristic regimes, which are separated by a time-dependent crossover momentum $q\sub{th}(\tau)$, which turns out to be the inverse thermal length scale $x\sub{th}(\tau)=1/q\sub{th}(\tau)$. According to the numerical simulations, $q\sub{th}(\tau)$ can be parametrized as $q\sub{th}(\tau)=Q_{\lambda}\tau^{\alpha_{\lambda}}$, where the exponent $\alpha_{\lambda}$ and the amplitude $Q_{\lambda}$ are monotonic functions of the quench parameter (for $\lambda>1$).  According to Fig.~\ref{fig:OccFull}, away from the crossover, the phonon distribution can be written as
\eq{Eq37}{
n_{\tau,q}=\left\{\begin{array}{cl}n_{\lambda}+c_{\tau,\lambda}|q|& \mbox{ for } |q|<q\sub{th}(\tau)\\ \frac{\tilde{T}_{\tau,\lambda}}{u |q|}& \mbox{ for } |q|>q\sub{th}(\tau)\end{array}\right. .
}
For small momenta $|q|<q\sub{th}$, the phonon density increases linearly in momentum, with a time-dependent prefactor $c_{\tau,\lambda}$, which has to be computed numerically but is determined solely by the quench parameter. This linear increase is guaranteed by the structure of the cubic vertex, which induces a scaling of the one-loop diagrams $\sim |q|$ for small momenta $q$. This scaling is imposed by the $U(1)$-symmetry of the action, which forbids a smaller exponent in the scaling of the local vertex as discussed in Ref.~\cite{buchholdmethod}, where  the same scaling behavior was found although with a different amplitude $c_{\tau}$ reflecting the driven nature of the system in that case. The very same mechanism guarantees the pinning of the distribution at $q=0$ to its initial value $n_{t,q=0}=n_{t=0,q=0}$, expressed by the constant $n_{\lambda}$ in Eq.~\eqref{Eq37}.\\
For larger momenta $|q|>q\sub{th}$ fast quasi-particle scattering events have established a local equilibrium and the phonon density is well described by a Bose distribution function $n\sub{B}(u|q|,\tilde{T}_{\tau,\lambda})=\left(e^{u|q|/\tilde{T}_{\tau,\lambda}}-1\right)^{-1}$, which can be approximated by a classical Rayleigh-Jeans distribution, as in Eq.~\eqref{Eq37}, for intermediate momenta $q\sub{th}<q<\frac{\tilde{T}_{\tau,\lambda}}{u}$. The effective temperature $\tilde{T}_{\tau,\lambda}$ approaches the final temperature $T_{\lambda}=\tilde{T}_{\tau\rightarrow\infty,\lambda}$ asymptotically, following a power law $\tilde{T}_{\tau,\lambda}-T_{\lambda}\sim \tau^{\mu}$ consistent with $\mu=2/3$ for large times, see section Fig.~\ref{sec:AsymTh}. 
An important exception is represented by the $q=0$ mode, which does not thermalize. The thermal momentum scale $q\sub{th}\sim \tau^{-\alpha_{\lambda}}$ is larger than zero for any finite time $\tau<\infty$. As a consequence, for any realistic experiment, there will always exist a small momentum window $q\in [0,q\sub{th}(t\sub{max})]$, which does not thermalize during the runtime of the experiment $t\sub{max}$. However, even in the limit $\tau\rightarrow\infty$, $n_{\tau,q=0}$ is pinned to its initial value by the exact $U(1)$-symmetry of the fermionic system. This symmetry corresponds to the exact particle number conservation of the fermionic theory. The occupation $n_{\tau,q=0}$ of the zero momentum mode is directly related to the variance of the total particle number $n_{\tau,q=0}\sim \langle \hat{N}^2\rangle_{\tau}-\langle \hat{N}\rangle^2_{\tau}$, where $\hat{N}$ is the total fermionic number operator \cite{Heating, mahan}. For particle number conserving dynamics, it is therefore an integral of motion. Consequently, the asymptotic bosonic distribution function in the limit $\tau\rightarrow\infty$ is a perfect Bose-Einstein distribution, with a discontinuity at $q=0$.

In order to express the factor $c_{\tau,\lambda}$ in terms of the temperature $\tilde{T}_{\tau,\lambda}$, we equate both forms of the distribution function $n_{\tau,q}$ in Eq.~\eqref{Eq37} at the crossover scale $q=q\sub{th}$. This yields an estimate for the non-equilibrium prefactor
\eq{Crosso}{c_{\tau,\lambda}=\frac{\tilde{T}_{\tau,\lambda}}{u q\sub{th}^2}.}

The off-diagonal densities $m_{\tau,q}$ are decaying in the long time limit, with $m_{\tau,q}\rightarrow0$ in the limit $\tau\rightarrow\infty$. Their time evolution is shown in Fig.~\ref{fig:OccFull} for the same quench parameters as used for the diagonal densities. For momenta larger than the crossover scale $q\sub{th}$, the off-diagonal densities decay exponentially fast in momentum, while they remain close to their initial value $m_{\lambda}$ for momenta smaller than the crossover. The number of scattering events into the off-diagonal modes is $\propto\Gamma^K_{\tau,q}$, which decreases in time very fast $\sim m_{\tau,q}^2$. This stands in contrast to the large number of out-scattering processes, which are given by $\sigma^R_{\tau,q}m_{\tau,q}\sim n_{\tau,q}m_{\tau,q}$ and dominate over the ingoing scattering events for a thermalizing diagonal distribution.

\subsection{Fermion Green's function}\label{sec:fac}
\begin{figure}
\begin{center}
  \includegraphics[width=1\linewidth]{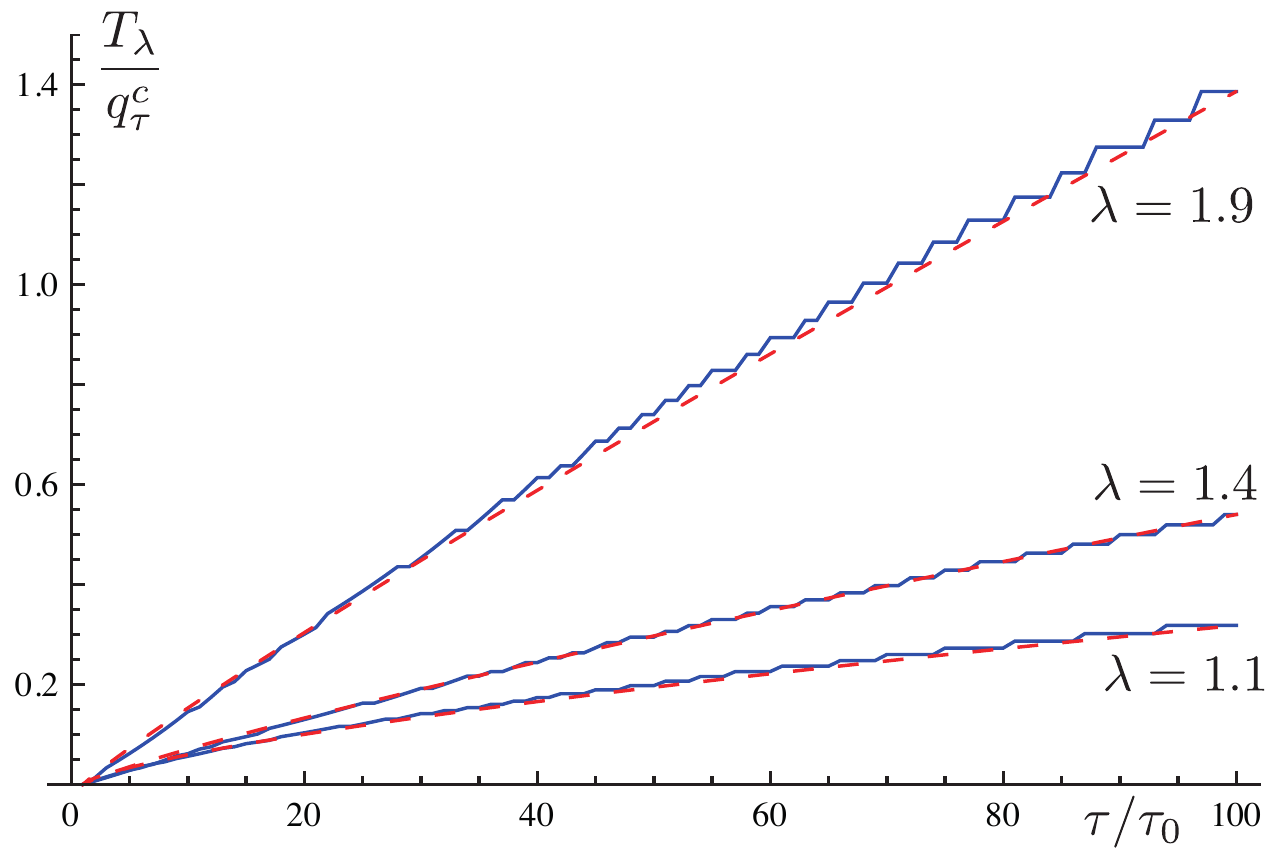}
  \caption{Time evolution of the thermal length scale $T_{\lambda}x\sub{th}(\tau)=\frac{T_{\lambda}}{q\sub{th}(\tau)}$
for three different quench scenarios $\lambda= (1.1, 1.4, 1.9)$, normalized with the corresponding
final temperature $T_{\lambda}$. The thermal length scale evolves according to a power law in time $T_{\lambda}x\sub{th}(\tau)=x_{\lambda}\tau^{\alpha_{\lambda}}$, where both amplitude and exponent depend non-trivially on the quench parameter and the exponent is invariant under a basis transformation from the dimensionless basis to the microscopic basis.
In the present example,  $\alpha_{1.1}= 0.7, \alpha_{1.4} = 0.85$ and $\alpha_{1.9} = 0.92$. The exponent is bounded from above $\alpha_{\lambda}<1$ due to the subleading nature of the interactions and from below $0<\alpha_{\lambda}$ by stability properties. For exponents $\alpha_{\lambda}<1$, the thermalization dynamics will always feature a finite spatio-temporal prethermalized region as indicated in Fig.~\ref{fig:QuenchDiag}.
 }
  \label{fig:Cross}
\end{center}
\end{figure}
The fermionic lesser Green's function $G^<_{t,x}$ can be computed using the time-evolved densities according to Eq.~\eqref{GF5}. The numerically determined fermion Green's function for a quench scenario with $\lambda=1.6$ are shown in Fig.~\ref{fig:Cross}, as discussed in the beginning of the section. One can identify three spatio-temporal regimes, with individually different, generic scaling behavior described by Eqs.~\eqref{eq:factorization2}-\eqref{tlength}. By exploiting the generic form of the time-evolved phonon densities for interacting Luttinger Liquids, we will in the following derive the form of the fermionic Green's function as given in Eq.~\eqref{eq:factorization2}.

In order to approximate the contribution from the off-diagonal densities, we exploit the fact that they remain close to their initial value $m_{\tau,q}\approx m_{\lambda}$ for momenta smaller than the crossover $q<q\sub{th}$ and decay exponentially for larger momenta, yielding a negligible influence on short distances. To account for this behavior, we replace in the corresponding integrals the cutoff $\Lambda\rightarrow q\sub{th}$ by the thermal crossover and approximate $m_{\tau,q}\approx m_{\lambda}$ for small momenta. The result is
\begin{eqnarray}
\mathcal{G}^<_{t,x}&=&\mathcal{G}^<_{0,x}+im_{\lambda}\log\left(\tfrac{1+q\sub{th}^2(x-2ut)^2}{1+q\sub{th}^2(x+2ut)^2}\right)\label{Eq38a}\\
&&-i\tfrac{(K^2-1)m_{\lambda}}{2K}\log\left[\tfrac{(1+q\sub{th}^2(x+2ut)^2)(1+q\sub{th}^2(x-2ut)^2)}{(1+4u^2t^2q\sub{th}^2)^2(1+x^2q\sub{th}^2)^2}\right]\label{Eq38b}\\
&&+i\tfrac{K^2+1}{K}\int_q\frac{2\pi e^{-\frac{|q|}{\Lambda}}}{|q|}(\cos(qx)-1)(n_{t,q}-n_{\lambda}).\label{Eq38c}
\end{eqnarray}
In this expression, $\mathcal{G}_{0,x}$ contains again the initial post-quench exponent, the terms proportional to $m_{\lambda}$ represent the time-dependent contributions stemming from the off-diagonal densities, whereas the first term vanishes for distances away from the prethermal crossover $x\neq 2ut$ and the second term vanishes for distances $x<1/q\sub{th}(\tau)=x\sub{th}(\tau)$ smaller than the thermal length. The latter expresses the fact, that off-diagonal occupations vanish in the asymptotic thermal limit. The term in Eq.~\eqref{Eq38c} takes into account the deviation of the diagonal phonon occupation from the flat initial distribution. Applying Eq.~\eqref{Eq37} to \eqref{Eq38c} with a smooth crossover function $\sim \exp(-|q|/q\sub{th}), 1-\exp(-|q|/q\sub{th})$, respectively,  amounts to
\begin{figure}
  \includegraphics[width=1\linewidth]{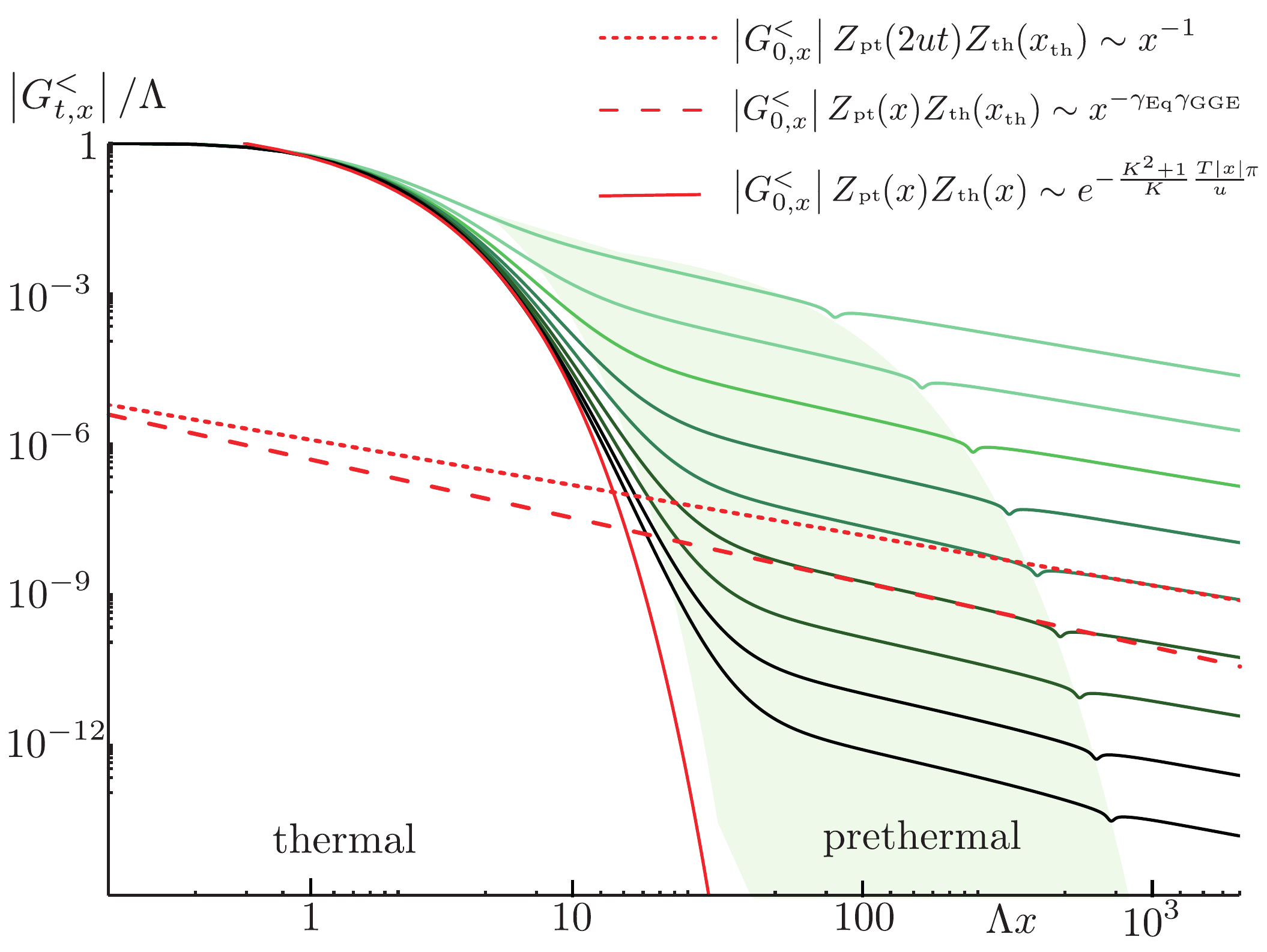}
  \caption{Numerical results for the absolute value of the fermionic lesser Green's function $G^<_{t,x}$ (green, solid lines) after a quench with $\lambda=1.6$ for different times $t=\frac{40}{u\Lambda}l$, $l=1,...,9$ (magnitude decreasing with $l$). The initial state corresponds to non-interacting fermions ($K=1$) and the interaction was chosen such that $v_0\Lambda^2=\frac{u\Lambda}{4}$. The figure illustrates the two crossovers and the three different spatio-temporal regimes of the Green's function. The red lines are determined according to the factorization \eqref{eq:factorization2}-\eqref{tlength} in the different regimes and describe the Green's function very well apart from the crossover regions.
}
  \label{fig:FermGreen}
\end{figure}
\begin{eqnarray}
\mathcal{G}^<_{t,x}&=&\mathcal{G}^<_{0,x}-i\tfrac{K^2-1}{2K}m_{\lambda}\log\left[\tfrac{(1+q\sub{th}^2(x+2ut)^2)(1+q\sub{th}^2(x-2ut)^2)}{(1+4u^2t^2q\sub{th}^2)^2(1+x^2q\sub{th}^2)^2}\right]\label{Eq39a}\\
&&-2\pi i\tfrac{K^2+1}{K}\tfrac{\tilde{T}_{\tau,\lambda}}{u}\left[\tfrac{q\sub{th}^2x^2}{1+q\sub{th}^2x^2}+|x|\left(1-\tfrac{2\arctan(q\sub{th}|x|)}{\pi}\right)\right],\label{Eq39b}
\end{eqnarray}
yielding the form of the fermionic Green's function in Eqs.~\eqref{eq:factorization2}-\eqref{tlength}. The prethermal amplitude $Z\sub{pt}$ is determined by the contribution $\sim m_{\lambda}$ in Eq.~\eqref{Eq39a}, while the thermal amplitude $Z\sub{th}$ is given by the exponential of the $\sim \tilde{T}_{\tau,\lambda}$-term in Eq.~\eqref{Eq39b}.
This form of the fermionic Green's function holds away from the crossover lines $|x|=2ut$ and $x=x\sub{th}(t)$. As shown in Fig.~\ref{fig:FermGreen}, it is a very good approximation for the fermionic Green's function and illustrates perfectly the different thermalization regimes and their scaling behavior.

\changed{The form of Eqs.~\eqref{Eq39a}-\eqref{Eq39b} allow us to estimate the distance $x_c$, below which the kinetic theory is applicable \mh{and which we have given already in \eqref{dist}}. One realizes immediately, that the prethermalization described by line \eqref{Eq39a} is absent for $K=1$, i.e. for a quench to the non-interacting theory. This is due to the absence of a coupling of the single particle sector to the many-body sector of the theory. A clear condition for the applicability of the kinetic theory can be obtained by comparing the time dependent variation of Eqs.~\eqref{Eq39a},\eqref{Eq39b} with each other. A well defined prethermal plateau has then been established for $|$\eqref{Eq39a}$|>||$\eqref{Eq39b}$|$, which leads to the condition on the distance $x<x_c(t)$, where $x_c(t)$ is given in Eq.~\eqref{dist}.}

We want to close the section by a discussion of the way, in which the microscopic scales enter the thermalization dynamics discussed in the present context. For the non-interacting Luttinger Liquid in equilibrium, the microscopic details are completely encoded in the sound velocity $u$ and the Luttinger parameter $K$ as well as the temperature of the system $T\ge 0$ and the Luttinger cutoff $\Lambda$. For a non-equilibrium setting in the quadratic Luttinger framework, one has to add the information on the initial state, which in the case of an interaction quench can be summarized in a single quench parameter $\lambda$. In the presence of interactions, we added the cubic vertex $\sim v_0$. This lead to the emergence of a new crossover scale $x\sub{th}(t)$, below which the system is effectively thermal, described by an effective time-dependent temperature $\tilde{T}_{t,\lambda}$. In the effective description of a factorizing Green's function, these quantities are sufficient to describe the post-quench dynamics. 


In the next section, we will discuss the time dependent temperature and find $\tilde{T}_{\lambda,\tau}=T_{\lambda}+\Delta_{\lambda}\tau^{-\mu}$, where $T_{\lambda}$ is the final temperature of the system, depending on the energy induced by the quench and $\Delta_{\lambda}$ the quench-dependent amplitude, while $\mu$ is a universal exponent. In original units, $\tilde{T}_{t,\lambda}=T_{\lambda}+u\Lambda\Delta_{\lambda}\left(v_0\Lambda^2t\right)^{-\mu}$. In the simplified picture, these are the only relevant quantities, which show a functional dependence on the nonlinearity $v_0$, naturally containing the limit $v_0\rightarrow0$, for which the thermal crossover is at zero distance and the temperature is not defined due to the absence of thermalization.

The thermalization dynamics for interacting Luttinger liquids presented so far is not restricted to interaction quenches or global quenches in general, but expected to represent quite generically the relaxation dynamics of Luttinger liquids out of equilibrium. First of all, the dephasing of the off-diagonal modes due to the quadratic Hamiltonian will be present in any setup for which off-diagonal modes have been excited in the initial state and it spreads in space with the light cone $x=2ut$. On the other hand, due to $U(1)$ symmetry and the imposed scaling of the one-loop correction $\sim q$ for small momenta $q$ \cite{buchholdmethod}, the change in the diagonal phonon distribution has to scale $\sim |q|$ as well. The determination of the crossover scale thus proceeds along the same lines as outlined above, and thus separating thermalized short distance modes with occupation $n_{t,q}\sim 1/|q|$ from non-thermal long distance modes $n_{t,q}-n_{0,q}\sim |q|$, leading to a similar three stage process for equilibration as described in the present setup.
\subsection{Asymptotic thermalization in the resonant approximation}\label{sec:AsymTh}
After the quench, momentum modes larger than the temporally decreasing crossover momentum $q\sub{th}$ establish a local detailed balance between in- and out-scattering processes. This in turn defines the thermalized region in real space, for which, on distances $x<x\sub{th}(t)=1/q\sub{th}(t)$, the fermionic Green's function has the typical thermal form. In this regime, the corresponding momentum modes for $q>q\sub{th}$ are described by a single, well defined temperature $\tilde{T}_{t,\lambda}$ such that $n_{t,q}=n_{\mbox{\tiny B}}(u|q|,T)\approx T/(u|q|)$. For momenta $q<q\sub{th}$ the phonon distribution is however smaller than the corresponding thermal distribution $u|q| n_{t,q}<T$ (see Fig.~\ref{fig:OccFull}) and in order to reach equipartition, energy has to be shifted from the thermal regime to the non-thermalized infrared modes. Consequently, the effective temperature of the large momentum modes is decreasing in time, expressing the energy flow from the large to the low momentum regime, i.e. 
 $\tilde{T}_{t\rightarrow\infty,\lambda}\rightarrow T_{\lambda}$ in the limit $x\sub{th}(t)\rightarrow\infty$.
The local equipartition of energy in the thermal regime is a consequence of a locally established detailed balance between in- and out-scattering processes. This local detailed balance in combination with exact energy conservation enforces that the energy transport to the long-wavelength modes in the system is performed by a global mechanism, which reveals the presence of dynamical slow modes in the system. They are a consequence of exact conservation laws, i.e. global symmetries, and in the present system emerge as a consequence of exact momentum and energy conservation. These modes are hydrodynamic, gapless modes featuring an algebraic decay of the temperature in time towards its final value.
\begin{figure*}
  \includegraphics[width=\linewidth]{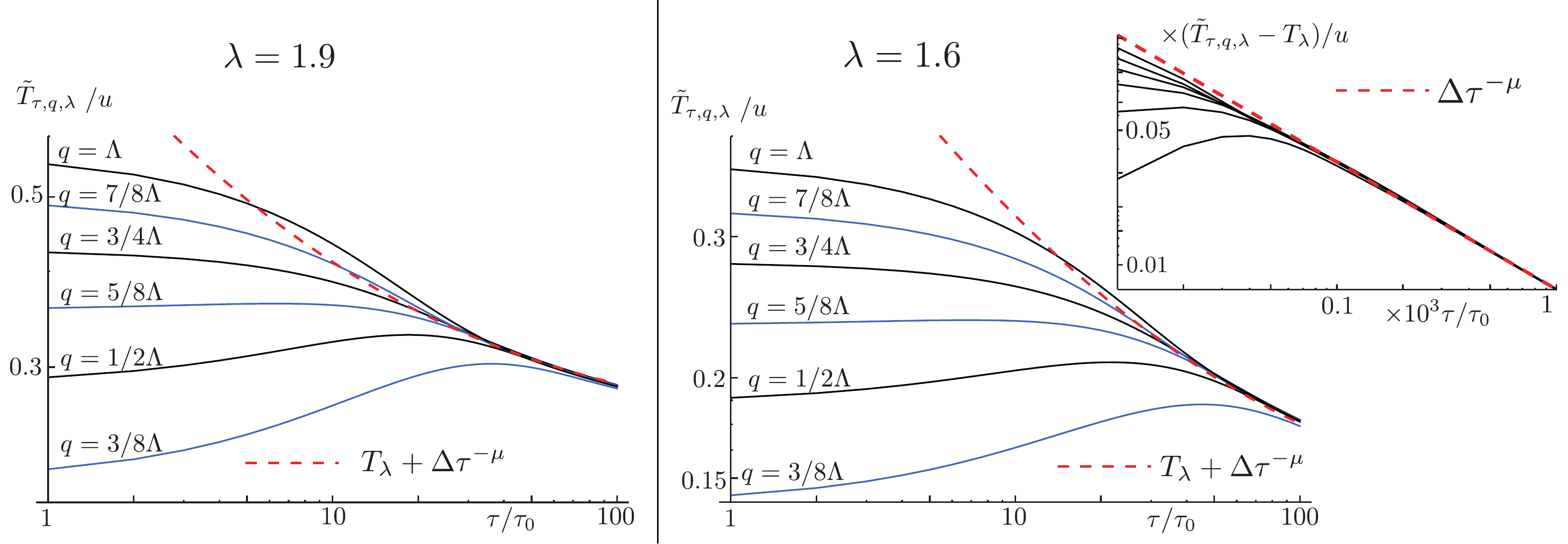}
  \caption{Momentum dependent temperature $\tilde{T}_{\tau,q,\lambda}$ as defined in Eq.~\eqref{Eq41} after two distinct quenches. For momenta smaller than the crossover $q\sub{th}$, the temperature is a momentum-dependent function and can not be seen as a global property. On the other hand, for momenta $q>q\sub{th}$, the modes are described by the same temperature, indicating the presence of local detailed balance in the momentum regime larger than the crossover. In this regime, the temperature decays algebraically, revealing energy transport from the thermalized to the non-thermalized region, carried by dynamical slow modes. The inset shows the decay of the effective temperature for large times, allowing for numerical estimate $\mu=2/3$, which corresponds to the red, dotted line.
}
  \label{fig:MomTherm}
\end{figure*}

In order to determine the asymptotic dynamics in the thermalized regime, we define a momentum and time dependent temperature by inverting the on-shell Bose distribution function 
\eq{Eq41}{
\tilde{T}_{t,\lambda,q}=\frac{u|q|}{\log\left(\frac{n_{t,q}+1}{n_{t,q}}\right).}
}
The time evolution of $\tilde{T}_{t,\lambda,q}$ is shown in Fig.~\ref{fig:MomTherm}. For momenta $q<q\sub{th}$ it varies as a function of momentum, indicating that the system has not thermalized on this scale and the notion of a temperature is absent. On the other hand, for momenta $q>q\sub{th}$, $\tilde{T}_{t,\lambda,q}$ becomes momentum independent and a global property of the high momentum modes. The decay of $\tilde{T}_{t,\lambda}=\tilde{T}_{t,q>q\sub{th},\lambda}$ follows a power law in time, which can be expressed
\eq{Eq42}{
\tilde{T}_{t,\lambda}=T_{\lambda}+u\Lambda\Delta_{\lambda}\left(v_0\Lambda^2t\right)^{-\mu},
}
where $\mu$ is the relaxation exponent associated with the dynamical slow modes. For a one-dimensional system with energy and momentum conserving dynamics $\mu=2/3$, since this behavior corresponds to the Kardar-Parisi-Zhang (KPZ) universality class \cite{KPZ,spohn04,lamacraft13,vanBeijeren,spohn15}. Performing a single parameter fit from the numerical simulations, we find that for large times $\mu=2/3$ agrees very well with the numerical data for various different quench scenarios. However, for intermediate times, we find scaling behavior with $\mu>2/3$ for some quenches, which might be traced back to the presence of subleading correction terms due to couplings to other diffusive modes \cite{Lux13,narayan02,Mukerjee}. Numerically a distinction of these possible scaling contributions is only possible for simulation times of multiple decades, such that we cannot exclude a different exponent $\mu<2/3$ at the largest times \cite{Lux13}, which is however not observed in our simulations.

While the establishment of a local detailed balance, leading to effective thermalization and thermal-like fermionic correlation functions is an effect of local quasi-particle scattering, the asymptotic thermalization dynamics describing energy transport over large distances in momentum space is determined by macroscopic diffusive modes in the system. This is observable by an algebraically decaying temperature towards the final temperature of the system $T_{\lambda}$. The discussion on the dynamical slow modes remains valid even in the presence of off-resonant scattering processes and therefore the universal properties of the asymptotic thermalization process remain unmodified. However, non-universal properties, such as the final temperature as well as the relaxation rate will be modified by the off-resonant processes. Their precise computation would be a task for numerical simulations.

\section{Conclusion}
In this work, we have analyzed the relaxation dynamics of interacting Luttinger liquids, microscopically represented by one-dimensional interacting fermions with band curvature, after a sudden quench in the fermionic interaction. The theoretical analysis is based on quantum kinetic equations for the phonon distribution function and non-perturbative Dyson-Schwinger equations, which are both well suited to determine the time-evolution of static observables for interacting Luttinger liquids with resonant, cubic interactions, and applicable in a broad parameter regime within the Luttinger framework. The central result is a two-step thermalization procedure including a spatio-temporal prethermalized regime for intermediate distances and times, which leads to fermionic correlation functions described by a generalized Gibbs state on these distances, and corresponds to fast quasi-particle formation after the quench. On smaller distances, a thermalized regime occurs due to the scattering and associated redistribution of energy between the quasi-particle modes. This regime is described by thermal correlation functions with a characteristic thermal correlation length and a thermal quasi-particle distribution with an effective temperature that decays algebraically in time towards its asymptotic value.

 This work shows in which way thermalization and prethermalization occur and spread in space for RG-irrelevant, and in this sense weak, integrability breaking interactions. In this setup both thermalization and prethermalization occur locally in space. While the prethermalized region spreads ballistically in space, the thermalized region spreads sub-ballistically due to the subleading, RG-irrelevant nature of the interactions. This allows for a well-defined prethermal regime in time and space, which would not be possible for a constant, momentum independent scattering vertex, for which thermalization would occur immediately on all different length scales.
This underpins the statement that typical candidates for clearly observable prethermalized regimes within generic thermalization dynamics are quasi-particle theories with RG irrelevant interactions.

\begin{acknowledgments}
We acknowledge valuable discussions with Alessio Recati. This research was supported by the
German Research Foundation (DFG) through the Institutional
Strategy of the University of Cologne within the German
Excellence Initiative (ZUK 81) and the European Research
Council (ERC) under the European Unions Horizon
2020 research and innovation programme (grant agreement
No 647434)
as well as the
Deutsche Akademie der Naturforscher Leopoldina under grant numbers LPDS 2013-07 and LPDR 2015-01.
\end{acknowledgments}

\bibliography{Diss.bib}

\appendix
\section{Keldysh action}\label{appendix2}
In this section, we derive the Keldysh action for the interacting Luttinger Liquid after the quench. The partition function as the generating functional of all possible correlation functions has the form
\eq{Ap1}{
Z(t)=\text{tr}\left(e^{-iHt}\rho_0e^{iHt}\right),
}
where $t$ is the time, $\rho_0$ is the initial state at $t=0$ and $H$ is the Hamiltonian \eqref{Eq5}, since we are interested in bosonic correlation functions. In order to express the partition function in terms of a path integral, one inserts bosonic coherent states at each infinitesimal time step and derives in a straight forward way the action on the $(\pm)$-contour,
\eq{Ap2}{
Z(t)=\int \mathcal{D}[\cre{a}{+,X},\cre{a}{-,X},\ann{a}{+,X},\ann{a}{-,X}]e^{i\mathcal{S}^{(\pm)}},
}
where $\mathcal{D}[...]$ is the common functional measure on the $(\pm)$-contour, $X=(x,t)$ the spatio-temporal coordinate and 
\eq{Ap3}{
\mathcal{S}^{(\pm)}&=&\int_{X}\sum_{\alpha=\pm}\alpha\Big(\cre{a}{\alpha,X}i\partial_t\ann{a}{\alpha,X}-H[\cre{a}{\alpha,X},\ann{a}{\alpha,X}]\Big)+\mathcal{F}
}
the action on the $(\pm)$-contour. In the action \eqref{Ap3}, the Hamiltonian is expressed in terms of $(\pm)$ fields by replacing the operators in Eq.~\eqref{Eq5} by complex fields. The functional $\mathcal{F}$ carries the information on the initial state (i.e. is the initial density matrix $\rho_0$ expressed in terms of the bosonic fields) and, depending on the precise initial stay, in general contains higher order vertices of arbitrary power \cite{Gutman10,Gutman10a}. After performing the Keldysh rotation to classical and quantum fields $\cre{a}{c}=(\cre{a}{+}+\cre{a}{-})/\sqrt{2}, \cre{a}{q}=(\cre{a}{+}-\cre{a}{-})/\sqrt{2}$ in the action, one obtains the Keldysh action\begin{widetext}
\begin{eqnarray}\label{Ap4}
\mathcal{S}&=&\int_{t,p}\left(\cre{a}{c,p,t},\cre{a}{q,p,t}\right)\left(\begin{array}{cc}0 & i\partial_t-u|q|-i0^+\\i\partial_t-u|q|+i0^+ & 0\end{array}\right)\left(\begin{array}{c}\ann{a}{c,p,t}\\ \ann{a}{q,p,t}\end{array}\right)\\
&&+\int_{t,p,k}'\hspace{-0.5cm}v_0\sqrt{|pk(p+k)|}\left[2\cre{a}{c,k+p,t}\ann{a}{c,p,t}\ann{a}{q,k,t}+\cre{a}{q,k+p,t}\left(\ann{a}{c,k,t}\ann{a}{c,p,t}+\ann{a}{q,k,t}\ann{a}{q,p,t}\right)+\mbox{H.c.}\right]\nonumber\\
&&+\mathcal{F}.\nonumber
\end{eqnarray}\end{widetext}
In this representation, the functional $\mathcal{F}$ contains only quantum fields \cite{Gutman10,Gutman10a,Chernii14} and, since it contains the information on the initial state, is uniquely determined by the complete set of irreducible correlation functions at $t=0$. In the present case, we consider an initial state, which is a thermal state corresponding to the prequench Hamiltonian and therefore the initial correlations correspond to thermal correlations, which, according to the Dzyaloshinkii-Larkin theorem \cite{Gutman10}, are only of quadratic order.
In the basis of the prequench fields, which we label as $\cre{b}{\alpha,p,t}, \ann{b}{\alpha,p,t}, \alpha=c,q$, $\mathcal{F}_{t=0}$ is therefore nothing but the thermal Keldysh self-energy
\eq{Ap5}{
\mathcal{F}_{t=0}=2i0^+\int_{p}\cre{b}{q,p,t=0}(2n(u|p|)+1)\ann{b}{q,p,t=0},
}
where $n(u|p|)$ is the Bose distribution.
The transformation from the prequench to the postquench basis can be performed by subsequently applying the canonical Bogoliubov transformation \eqref{Bog1}, \eqref{Bog2} and reads
\eq{Ap6}{
\cre{a}{\alpha,p,t}=\frac{1}{2}\left[\sqrt{\lambda}\left(\cre{b}{\alpha,p,t}-\ann{b}{\alpha,-p,t}\right)+\frac{1}{\sqrt{\lambda}}\left(\cre{b}{\alpha,p,t}+\ann{b}{\alpha,-p,t}\right)\right].}
Combining these results, the quantum part of the action can be expressed solely by the Keldysh self-energy,
\eq{Ap7}{
\mathcal{F}=\int_{p,t,t'}\left(\cre{a}{q,p,t},\ann{a}{q,-p,t}\right)\Sigma^K_{p,t,t'}\left(\begin{array}{c}\ann{a}{q,p,t'}\\ \cre{a}{q,-p,t'}\end{array}\right),
}with the initial condition
\eq{Ap8}{
\Sigma^K_{p,0,0}=\frac{2i0^+}{2\lambda}(2n(u|p|)+1)\left(\begin{array}{cc}1+\lambda^2 & \lambda^2-1\\ \lambda^2-1 & \lambda^2+1
\end{array}\right).}
The time evolution of the Keldysh self-energy and the corresponding phonon distribution function is determined via the kinetic equation approach in the main text.
\section{Fermionic Green's functions}
In this section, we derive the expression for the exponent \eqref{GF1} in the fermionic Green's function \eqref{GF4}. The fermionic lesser and greater Green's functions for right movers at equal times are defined as
\eq{Ap9}{
G^<_{t,x}=-i\langle \cre{\psi}{t,x}\ann{\psi}{t,0}\rangle=-i\langle \cre{\psi}{-,t,x}\ann{\psi}{+,t,0}\rangle,\\
G^>_{t,x}=-i\langle \ann{\psi}{t,x}\cre{\psi}{t,0}\rangle=i\langle \cre{\psi}{+,t,-x}\ann{\psi}{-,t,0}\rangle,
}
where the second equality in both equations indicates the average with respect to the functional integral and the indices $(\pm)$ denote the corresponding contour. The corresponding Green's functions for left movers are obtained by $x\rightarrow-x$, as discussed in the main text. Obviously, in a spatially translational invariant system, the greater Green's function is obtained from the lesser Green's function by a contour exchange ($+\leftrightarrow-$) and spatial inversion ($x\rightarrow-x$). Right moving fermion operators are expressed in terms of Luttinger fields according to
\eq{Ap10}{
\cre{\psi}{\alpha,t,x}=\sqrt{\frac{\Lambda}{2\pi}}\ e^{ik\sub{F}x} e^{i\left(\phi_{\alpha,t,x}-\theta_{\alpha,t,x}\right)},}
where $\alpha=\pm$ labels the contour. It is important to perform the transformation \eqref{Ap10} on the $(\pm)$- and not on the Keldysh-contour since the transformation to the Luttinger basis does not commute with the Keldysh rotation. The lesser Green's function expressed in terms of the Luttinger fields is
\eq{Ap11}{
G^<_{t,x}=-i\frac{\Lambda}{2\pi}e^{ik\sub{F}x}\langle e^{i(\phi_{-,t,x}-\theta_{-,t,x}-\phi_{+,t,0}+\theta_{+,t,0})}\rangle=-i\frac{\Lambda}{2\pi}e^{ik\sub{F}x} e^{-\frac{i}{2}\mathcal{G}^<_{t,x}}.}
The exponent
\eq{Ap12}{
\mathcal{G}^<_{t,x}=2i\log \langle e^{i(\phi_{-,t,x}-\theta_{-,t,x}-\phi_{+,t,0}+\theta_{+,t,0})}\rangle
}
is according to the linked cluster theorem nothing but the sum of all one-particle irreducible contractions of an expansion of the exponential. The generating functional for the one-particle irreducible contractions is the effective action $\Gamma[\cre{a}{\alpha,X},\ann{a}{\alpha,X}]$, which we determine up to cubic order by Dyson-Schwinger equations. The four-point irreducible vertex is subleading an negligibly small on all relevant scales. On the other hand, the three-body irreducible vertex remains local \cite{buchholdmethod} and therefore in the expansion of the exponential \eqref{Ap11}, only purely local terms (e.g. $\sim \theta_{t,x}^3$) give a contribution at cubic order. Due to translational invariance, these contributions yield only a constant amplitude for the Green's function, which due to of the Green's function must be unity. Consequently, only the quadratic terms contribute to the expansion and the sum of all quadratic irreducible vertices is the full Green's function, i.e.
\eq{Bp6}{
\mathcal{G}^<_{t,x}=-i\langle (\phi_{-,t,x}-\theta_{-,t,x}-\phi_{+,t,0}+\theta_{+,t,0})^2\rangle.}
Performing the Keldysh rotation, one straightforwardly arrives at the expression for the exponent \eqref{GF4}. 

The various Green's functions can be evaluated using the Bogoliubov transformation to the phonon basis, this yields the set of Green's functions
\begin{eqnarray}\label{Bp7a}
G^R_{\theta\theta,t,x}&=&G^R_{\phi\phi,x,t}=0,\\
G^K_{\phi\phi,t,x}&=&\int_q \frac{\pi K}{|q|} \left(G^K_{t,q}+i\mbox{Im}(G^{KA}_{t,q})\right)\cos(qx)e^{-\frac{|q|}{\Lambda}},\\
G^K_{\theta\theta,t,x}&=&\int_q \frac{\pi}{|q|K} \left(G^K_{t,q}-i\mbox{Im}(G^{KA}_{t,q})\right)\cos(qx)e^{-\frac{|q|}{\Lambda}},\ \ \ \ \ \  \ \ \ \\
G^R_{\theta\phi,t,x}-G^A_{\theta\phi,t,x}&=&\hspace{-0.15cm}-i\hspace{-0.1cm}\int_q \frac{\pi}{q}\sin(qx)\left(G^R_{t,q}-G^A_{t,q}\right)e^{-\frac{|q|}{\Lambda}}\hspace{-0.1cm}\nonumber\\
&=&-\arctan(\Lambda x),\\
G^K_{\theta\phi,t,x}&=&-i\int_q\frac{\pi}{q}\sin(qx)\mbox{Re}(G^{KA}_{t,q})e^{-\frac{|q|}{\Lambda}}.\end{eqnarray}
Here, $G^K_{t,q}=-i\langle \ann{a}{c,q,t}\cre{a}{c,q,t}\rangle=-i(2n_{t,q}+1)$ is the equal time diagonal Keldysh Green's function and $G^{KA}_{t,q}=-i\langle\cre{a}{c,-q,t}\cre{a}{c,q,t}\rangle=-i2m_{t,q}e^{2iu|q|t}$ is the anomalous equal time Keldysh Green's function.
Inserting these expressions in the exponent \eqref{GF4}, one finds \eqref{GF5}, which is $\mathcal{G}^<_{t,x}$ for equal times up to fourth oder irreducible vertex corrections.

\end{document}